\documentclass[12pt]{article}%
\usepackage{amsmath}
\usepackage{amsfonts}
\usepackage{amssymb}
\usepackage{graphicx}
\usepackage[utf8]{inputenc}%
\providecommand{\U}[1]{\protect\rule{.1in}{.1in}}
\newtheorem{theorem}{Theorem}
\newtheorem{acknowledgement}[theorem]{Acknowledgement}

\begin{document}
\begin{titlepage}
\vspace{.3cm} \vspace{1cm}
\begin{center}
\baselineskip=16pt \centerline{\Large\bf  Quanta of Space-Time and Axiomatization of Physics\footnote
{Contribution to the special issue of IJGMMP celebrating the one century anniversary of the program announced in 1916 by
Hilbert entitled {\it Foundations of Mathematics and Physics}, editors, Joseph Kouneiher, John Stachel and Salvatore
Capozzieolo} } \vspace{2truecm}
\centerline{\large\bf Ali H.
Chamseddine$^{1,2}$ } \vspace{.5truecm}
\emph{\centerline{$^{1}$Physics Department, American University of Beirut, Lebanon}}
\emph{\centerline{$^{2}$I.H.E.S. F-91440 Bures-sur-Yvette, France}}
\end{center}
\vspace{2cm}
\begin{center}
{\bf Abstract}
\end{center}
We consider Hilbert's sixth problem on the axiomatization of physics starting with
a higher degree Heisenberg commutation relation involving the Dirac operator
and the Feynman slash of scalar fields. The two sided version of the commutation relation in dimension $4$ implies
volume quantization and determines a noncommutative space which is a tensor  product of continuous and discrete
spaces. This noncommutative space predicts the full structure of a unified model of all particle interactions based on
Pati-Salam symmetries or, as a special case, the Standard Model.  We study implications of this
quantization condition on Particle Physics, General Relativity, the cosmological constant and dark matter.
We demonstrate that, with little input, noncommutative geometry gives a compelling and attractive  picture about the
nature and structure of space-time.
\end{titlepage}

\section{Introduction}

\bigskip David Hilbert research on the axiomatization of geometry led him to
suggest the sixth problem on his list for the axiomatization of Physics which
have received the least attention \cite{Corey}. Hilbert contributed
prominently to the formulation of the gravitational equations in the General
Theory of Relativity which was presented in November 1915, almost
simultaneously with Einstein \cite{Hilbert1} \cite{Hilbert2}. Weyl has
asserted that during the period 1910-1922 Hilbert has devoted considerable
time to research in Physics which was an integral part of his mathematical
world. Indeed, in 1915 Hilbert has presented a unified theory of
electromagnetism and gravitation based on the use of the variational principle
derived in an axiomatic fashion from the two principles of general invariance
and "Mie's axiom of the world function". This attempt can be considered as the
seed that motivated much work on ideas on unification of all fundamental
interactions such as in Kaluza-Klein theory, supersymmetry, superstring theory
and noncommutative geometry. In this article I will follow up on the
contribution of Alain Connes to this volume and show that starting with the
axioms of noncommutative geometry supplemented by a minimal number of physical
assumptions would result, unambiguously, in a unified theory of all
fundamental interactions and matter content of space-time \cite{Connesbook},
\cite{Gracia}. We will be able to establish a link between the quantization of
volume of space at Planck energy and the constituents of matter and their
symmetries. In addition we uncover the origin of the Higgs fields and symmetry
breaking, and indicate possible solutions to long standing problems such as
resolving the singularities in GR, dark matter and dark energy.

All the material covered in this review is a result of a long time
collaboration with Alain Connes which started in 1996 and continues until now.
More recently our collaboration included Walter van Suijlekom and, in separate
publications, Slava Mukhanov. An excellent introduction to the material
covered in this review is the accompanying article by Alain Connes in this
volume. However, an attempt is made to make this article self-contained.

The Planck scale is the scale at which all rescaled curvature invariants of a
Riemannian manifold are of the same order. The volume of any manifold at
scales below the Planck scale, will be many orders of magnitude larger than
that scale. To avoid the problem of infinities, which are expected to arise in
a quantized theory of gravity, it is a natural proposition to assume that the
volume of a physical space is an integer multiple of a unit volume of
Planckian size and thus provide a cutoff scale. It is well known that the
degree of a smooth map $Y$ from a connected, compact, oriented $n$-manifold to
the sphere $S^{n}$ is an integer
\begin{equation}
Y:M_{n}\rightarrow S^{n},
\end{equation}
where $Y$ is $\mathbb{R}^{n+1}$ valued on $M_{n}.$ This map is normalized by
$\left\langle Y\left(  x\right)  ,Y\left(  x\right)  \right\rangle =1$ where
$x\in M_{n}$ and if we let $\Delta$ be the positive normed determinant
function in $\mathbb{R}^{n+1}$, then the degree of the map is given by
\cite{Greub}
\begin{equation}
\deg\left(  Y\right)  \equiv\frac{1}{\kappa_{n}}%
%TCIMACRO{\dint \limits_{M_{n}}}%
%BeginExpansion
{\displaystyle\int\limits_{M_{n}}}
%EndExpansion
\left\langle \Delta,Y\left(  dY\right)  ^{n}\right\rangle \in\mathbb{Z}
\label{winding}%
\end{equation}
where $\kappa^{n}$ is the volume of the $n$-sphere:
\begin{equation}
\kappa_{2m}=\frac{2^{m+1}}{\left(  2m-1\right)  !!}\pi^{m},\qquad\kappa
_{2m+1}=\frac{2}{m!}\pi^{m+1},\quad m=1,\cdots,\infty.
\end{equation}
We propose to identify the integrand in (\ref{winding}), which is an $n$-form
over an $n$-dimensional connected, compact oriented manifold, with the volume
form:
\begin{equation}
w_{n}=\frac{1}{\kappa_{n}}\left\langle \Delta,Y\left(  dY\right)
^{n}\right\rangle , \label{heisenberg}%
\end{equation}
then the volume of $M_{n}$ will be an integer multiple of the unit Planckian
$n$-sphere. From this we deduce that the pullback $Y^{\ast}\left(
w_{n}\right)  $ is a differential form that does not vanish anywhere. This in
turn implies that the Jacobian of the map $Y$ does not vanish anywhere, and
that $Y$ is a covering of the sphere. The sphere is simply connected, and on
each connected component $M_{j}\subset M_{n}$, the restriction of the map $Y$
to $M_{j}$ is a diffeomorphism, implying that the manifold must be
disconnected, with each piece having the topology of a sphere \cite{CCM1}. We
will show how to avoid this unsatisfactory conclusion and how the attractive
idea of volume quantization works in a convincing way within the formulation
of noncommutative geometry.

Extensive research over the last two decades have shown that there are many
advantages to work with noncommutative geometry instead of Riemannian geometry
\cite{Connesbook}. The approach is spectral in nature and its concepts are
modeled after quantum mechanics where geometry is defined in terms of spectral
data. These are specified in terms of spectral triple $\left(  \mathcal{A}%
,\mathcal{H},D\right)  $ where $\mathcal{A}$ is an associative algebra with
unit $1$ and involution $\ast$, $\mathcal{H}$ a complex Hilbert space carrying
a faithful representation of the algebra $\mathcal{A}$ and $D$ is a
slef-adjoint operator on $\mathcal{H}$ with the resolvent $\left(
D-\lambda1\right)  ^{-1},$ where $\lambda\notin\mathbb{R}$ of $D,$ compact.
The operator $D$ plays the role of inverse line element. In addition the real
structure $J$ is an anti-unitary operator that sends the algebra $\mathcal{A}$
to its commutant $\mathcal{A}^{o}$ such that \cite{Reality}
\begin{equation}
\left[  a,b^{o}\right]  =0,\qquad a,b\in\mathcal{A},\qquad b^{o}=Jb^{\ast
}J^{-1}\in\mathcal{A}^{o}.
\end{equation}
The chirality operator $\gamma$ is a unitary operator in $\mathcal{H}$ defined
in even dimensions such that $\gamma^{2}=1$ and commutes with $\mathcal{A}$
\begin{equation}
\left[  \gamma,a\right]  =0\qquad\forall a\in\mathcal{A}.
\end{equation}
There are commutativity or anti-commutativity relations between $D,$ $J,$ and
$\gamma:$%
\begin{equation}
J^{2}=\epsilon,\qquad JD=\epsilon^{\prime}DJ,\qquad J\gamma=\epsilon
^{\prime\prime}\gamma J,\qquad D\gamma=-\gamma D,
\end{equation}
where $\epsilon,\epsilon^{\prime},\epsilon^{\prime\prime}\in\left\{
-1,1\right\}  .$ The operators $\gamma$ and $J$ are similar to the chirality
and charge conjugation operators and to every fixed value of $\epsilon
,\epsilon^{\prime},\epsilon^{\prime\prime}$ is associated a KO dimension,
which may be non-metric, and thus is defined only modulo $8.$ It is then
evident that the generalized Heisenberg relation must be modified to include
not only the mapping $Y$ from $M_{n}$ to $S^{n}$ but also the effects of the
operator $J$ which requires two mappings $Y$ and $Y^{\prime}.$ We have shown
that using the two mappings $Y$ and $Y^{\prime}$ to set the volume
quantization condition would avoid limiting the topology of the manifold to be
that of a sphere in dimensions two and four \cite{CCM} \cite{CCM1}. We shall
elaborate on the form of the generalized Heisenberg relation and show that
this leads, unambiguously, to the construction of a noncommutative space whose
geometry gives naturally a unified model of all particle interactions based on
Pati-Salam symmetry group which also includes the Standard Model as a special
case .

This article is organized as follows. In section 2 the conjectured Heisenberg
quantization two sided relation is constructed in such a way as to give the
volume of the underlying manifold to be given by the sum of two integers times
the volume of a unit Planckian sphere. In section three the algebra of the
finite noncommutative space is derived to be the sum of two algebras, which in
dimension four, is given by the sum $M_{2}\left(  \mathbb{H}\right)  $ and
$M_{4}\left(  \mathbb{C}\right)  $ \cite{CC2}, \cite{CCS} \cite{CCS2}. In
section four we determine the noncommutative space and make contact with our
previous work on noncommutative geometry \cite{AC2}, \cite{saddam}, \cite{AC}.
In section five we show the the unified model associated with this
noncommutative space is of the Pati-Salam type and in section six we give the
Standard Model obtained as a limiting case \cite{AC}. Section seven is a
summary of the minimal Pati-Salam model \cite{CCS2}, \cite{Walter}. In section
8 we present the spectral action principle and calculate the spectral action
of the Standard Model. In section 9 we study consequences of volume
quantization on the equations of motion in both instances when the fields $Y$
and $Y^{\prime}$ are with or without kinetic terms. In section 10 we give the
solitonic solutions and show that these are identical to the $O(5)$ non-linear
gravitational sigma model. In section 11 we consider the case of a Riemannian
manifold with Lorentzian signature where the four-dimensional manifold is
viewed as a $3+1$ space formed from the motion of three dimensional
hypersurfaces. We show that it is possible to impose quantization of the three
dimensional compact space provided that the field mapping the one-dimensional
non-compact space satisfies a length preserving relation. In section 12 we
further discuss the conditions under which a quantization of a two dimensional
hypersurface is possible. In section 13 we study the equations of motion for
the cases of three dimensional volume and two dimensional surface
quantization. In section 14 we discuss quantization on the special spaces
$\mathbb{R}\times S^{3}$ and $\mathbb{R}^{2}\times S^{2}.$ Section 15 contains
a discussion and the conclusion.

\section{Heisenberg volume quantization in dimensions 2 and 4}

For a Riemannian manifold of dimension $n$ the algebra $\mathcal{A}$ is taken
to be $C^{\infty}\left(  M\right)  ,$ the algebra of continuously
differentiable functions, while the operator $D$ is identified with the Dirac
operator given by
\begin{equation}
D_{M}=\gamma^{\mu}\left(  \frac{\partial}{\partial x^{\mu}}+\omega_{\mu
}\right)  ,
\end{equation}
where $\gamma^{\mu}=e_{a}^{\mu}\gamma^{a}$ and $\omega_{\mu}=\frac{1}{4}%
\omega_{\mu bc}\gamma^{bc}$ is the $SO(n)$ Lie-algebra valued spin-connection
with the (inverse) vielbein $e_{a}^{\mu}$ being the square root of the
(inverse) metric $g^{\mu\nu}=e_{a}^{\mu}\delta^{ab}e_{b}^{\nu}.$ The gamma
matrices $\gamma^{a}$ are anti-hermitian $\left(  \gamma^{a}\right)  ^{\ast
}=-\gamma^{a}$ and define the Clifford algebra $\left\{  \gamma^{a},\gamma
^{b}\right\}  =-2\delta^{ab}.$ The Hilbert space $\mathcal{H}$ is the space of
square integrable spinors $L^{2}\left(  M,S\right)  .$ The Dirac operator is
Hermitian with respect to the inner product
\begin{equation}
\left(  \psi,D_{M}\psi\right)  =\left(  D_{M}\psi,\psi\right)  =%
%TCIMACRO{\dint }%
%BeginExpansion
{\displaystyle\int}
%EndExpansion
d^{n}xe\psi^{\ast}D_{M}\psi,
\end{equation}
where $e=\det\left(  e_{\mu}^{a}\right)  $ with $e_{\mu}^{a}$ being the
inverse of $e_{a}^{\mu}.$ The chirality operator $\gamma$ in even dimensions
is then given by
\begin{equation}
\gamma=\left(  i\right)  ^{\frac{n}{2}}\gamma^{1}\gamma^{2}\cdots\gamma^{n}%
\end{equation}
From the above discussion, it is very suggestive to associate with the map
fields $Y^{A},$ $A=1,2,\cdots,n+1$ a Clifford algebra valued field
$Y=Y^{A}\Gamma_{A}$ where \cite{LM}
\begin{equation}
\Gamma_{A}\in C_{\kappa},\quad\left\{  \Gamma_{A},\Gamma_{B}\right\}
=2\kappa\,\delta_{AB},\ (\Gamma_{A})^{\ast}=\kappa\Gamma_{A}.
\end{equation}
Here $\kappa=\pm1$ and $C_{\kappa}\subset M_{s}(\mathbb{C})$ is the algebra of
$s\times s$ matrices, where $s=2^{n/2}.$ A\ generalization of the Heisenberg
commutation relation $[p,q]=-i\hbar$ is conjectured to be given by \cite{CCM1}%
\begin{equation}
\left\langle Y\left[  D,Y\right]  \cdots\left[  D,Y\right]  \right\rangle
=\sqrt{\kappa}\,\gamma\quad\left(  n\mathrm{\ terms\,}\left[  D,Y\right]
\right)  , \label{yyy}%
\end{equation}
where $Y\in C^{\infty}\left(  M\right)  \otimes C_{\kappa}$ is of the Feynman
slashed form $Y=Y^{A}\Gamma_{A},$ and fulfill the equations
\begin{equation}
Y^{2}=\kappa,\qquad Y^{\ast}=\kappa Y.
\end{equation}
The notation $\left\langle T\right\rangle $ means the trace of $T$ with
respect to the above matrix algebra $M_{s}(\mathbb{C}).$ In a coordinate basis
equation (\ref{yyy}) takes the form \cite{CCM1}%
\begin{equation}
\frac{1}{n!}\epsilon^{\mu_{1}\mu_{2}\cdots\mu_{n}}\epsilon_{A_{1}A_{2}\cdots
A_{n+1}}Y^{A_{n+1}}\partial_{\mu_{1}}Y^{A_{1}}\partial_{\mu_{2}}Y^{A_{2}%
}\ldots\partial_{\mu_{n}}Y^{A_{n}}=\det\left(  e_{\mu}^{a}\right)  ,
\end{equation}
which is a constraint on the volume form. This can be thought of as a
generalization of the coordinate-momenta $\left[  p,q\right]  =-i\hbar$ phase
space quantization where $p$ is replaced with the Dirac operators $D$ and $q$
is replaced with the Feynman slash coordinates $Y$. We have seen, however,
that this quantization condition implies that the $n-$manifold decomposes into
a set of bubbles. The difference now is that the quantization condition is
given in terms of the noncommutative data. One cannot fail to notice that the
operator $J$ is missing from equation (\ref{yyy}) which suggests that this
equation must be modified to take this operator into account. We first define
the projection operator $e=\frac{1}{2}\left(  1+Y\right)  $ satisfying
$e^{2}=e$ \ \cite{Connes} but now there are two possibilities, $Y$
corresponding to the case $\kappa=1$ and $Y^{\prime}$ to the case $\kappa=-1$.
Thus let $Y=Y^{A}\Gamma_{A}\equiv Y$ and let $Y^{\prime}=iJYJ^{-1}$ and
$\Gamma_{A}^{\prime}=iJ\Gamma_{A}J^{-1}$ so that we can write
\begin{equation}
Y=Y^{A}\Gamma_{A},\qquad Y^{\prime}=Y^{\prime A}\Gamma_{A}^{\prime},
\end{equation}
satisfying $Y^{2}=1$ and $Y^{\prime2}=1.$ The projection operators $e=\frac
{1}{2}\left(  1+Y\right)  $ and $e^{\prime}=\frac{1}{2}\left(  1+Y^{\prime
}\right)  $ satisfy $e^{2}=e$, $e^{\prime2}=e^{\prime}$ with $e$ and
$e^{\prime}$ commuting. This allows to define the projection operator
$E=ee^{\prime}$ and the associated field
\begin{equation}
Z=2E-1,
\end{equation}
satisfying $Z^{2}=1.$ The conjectured quantization condition takes the elegant
form of a two-sided relation \cite{CCM1}, \cite{CCM}%

\begin{equation}
\left\langle Z\left[  D,Z\right]  ^{n}\ \right\rangle =\gamma. \label{SM}%
\end{equation}
Our proposal is that this quantization condition is valid for all
noncommutative geometries defined by the spectral data where the metric
dimension of the operator $D$ as determined from the Weyl asymptotic formula
is less than or equal to four. The presence of the chirality operator $\gamma$
indicates that the dimension $n$ should be even, and this would limit us to
the two cases $n=2$ and $n=4.$ For odd dimensional $n$ the form of the
quantization condition should be modified, but will not be considered here. We
have shown that for both $n=2$ and $n=4$ equation (\ref{SM}) splits as the sum
of two pieces \cite{CCM1}%
\begin{equation}
\left\langle Z\left[  D,Z\right]  ^{n}\right\rangle =\left\langle Y\left[
D,Y\right]  ^{n}\right\rangle +\left\langle Y^{\prime}\left[  D,Y^{\prime
}\right]  ^{n}\right\rangle .
\end{equation}
This implies that the volume form of the $n-$dimensional Riemannian manifold
is the sum of two $n-$forms and thus
\begin{align}
\det\left(  e_{\mu}^{a}\right)   &  =\frac{1}{n!}\epsilon^{\mu_{1}\mu
_{2}\cdots\mu_{n}}\epsilon_{A_{1}A_{2}\cdots A_{n+1}}Y^{A_{n+1}}\partial
_{\mu_{1}}Y^{A_{1}}\partial_{\mu_{2}}Y^{A_{2}}\ldots\partial_{\mu_{n}}%
Y^{A_{n}}+\\
&  +\frac{1}{n!}\epsilon^{\mu_{1}\mu_{2}\cdots\mu_{n}}\epsilon_{A_{1}%
A_{2}\cdots A_{n+1}}Y^{^{\prime}A_{n+1}}\partial_{\mu_{1}}Y^{^{\prime}A_{1}%
}\partial_{\mu_{2}}Y^{A_{2}}\ldots\partial_{\mu_{n}}Y^{\prime A_{n}}.
\end{align}
Consider the smooth maps $\phi_{\pm}:M_{n}\rightarrow S^{n}$ then their
pullbacks $\phi_{\pm}^{\#}$ would satisfy
\begin{equation}
\phi_{+}^{\#}\left(  \alpha\right)  +\phi_{-}^{\#}\left(  \alpha\right)
=\omega, \label{integer}%
\end{equation}
where $\alpha$ is the volume form on the unit sphere $S^{n}$ \cite{Moser} and
$\omega\left(  x\right)  $ is an $n-$form that does not vanish anywhere on
$M_{n}.$ We stress that the quantization condition does not split as the sum
of two terms except for $n=2,4$, however, if one starts with the conjecture
that the volume form is the sum of the two traces in terms of the coordinates
$Y$ and $Y^{\prime}$ then equation (\ref{integer}) would follow and would then
not be limited to the two values for $n.$ We have shown that for a compact
connected smooth oriented manifold with $n<4$ one can find two maps $\phi
_{+}^{\#}\left(  \alpha\right)  $ and $\phi_{-}^{\#}\left(  \alpha\right)  $
whose sum does not vanish anywhere, satisfying equation (\ref{integer}) such
that $%
%TCIMACRO{\dint \limits_{M}}%
%BeginExpansion
{\displaystyle\int\limits_{M}}
%EndExpansion
\omega\in\mathbb{Z}.$ The proof for $n=4$ is more difficult and there is an
obstruction unless the second Stieffel-Whitney class $w_{2}$ vanishes, which
is satisfied if $M$ is required to be a spin-manifold and the volume to be
larger than or equal to five units. The key idea in the proof is to note that
the kernel of the map $Y$ is a hypersurface $\Sigma$ of co-dimension $2$ and
therefore \cite{CCM1}
\begin{equation}
\dim\Sigma=n-2.
\end{equation}
We can then construct a map $Y^{\prime}=Y\circ\psi$ where $\psi$ is a
diffeomorphism on $M$ such that the sum of the pullbacks of $Y$ and
$Y^{\prime}$ does not vanish anywhere. The important point to stress here is
that the conjectured two sided relation (\ref{SM}) is taken to hold for
arbitrary noncommutative spaces where $n\leq4$ where $n$ \ is the dimension as
determined in the Weyl asymptotic formula for the growth of eigenvalues of the
Dirac operator, and is not restricted for Riemannian manifolds. In other
words, one can seek solutions for this equation in general and find the
noncommutative space satisfying this equation.

\section{Clifford Algebras and Feynman slash}

We have seen that the coordinates $Y$ are defined over a Clifford algebra
$C_{+}$ spanned by $\left\{  \Gamma_{A},\Gamma_{B}\right\}  =2\delta_{AB}.$
For $n=2$, $C_{+}=M_{2}\left(  \mathbb{C}\right)  $ while for $n=4$,
$C_{+}=M_{2}\left(  \mathbb{H}\right)  \oplus M_{2}\left(  \mathbb{H}\right)
$ where $\mathbb{H}$ is the field of quaternions \cite{LM}. However, for
$n=4,$ since we will be dealing with irreducible representations we take
$C_{+}=M_{2}\left(  \mathbb{H}\right)  .$ Similarly the coordinates
$Y^{\prime}$ are defined over the Clifford algebra $C_{-}$ spanned by
$\left\{  \Gamma_{A}^{\prime},\Gamma_{B}^{\prime}\right\}  =-2\delta_{AB}$ and
for $n=2$, $C_{-}=\mathbb{H\oplus H}$ and for $n=4$, $C_{-}=M_{4}\left(
\mathbb{C}\right)  .$ The operator $J$ acts on the two algebras $C_{+}\oplus
C_{-}$ in the form $J\left(  x,y\right)  =\left(  y^{\ast},x^{\ast}\right)  $
(i.e. it exchanges the two algebras and takes the Hermitian conjugate). The
coordinates $Z=\frac{1}{2}\left(  Y+1\right)  \left(  Y^{\prime}+1\right)
-1,$ then define the matrix algebras \cite{CC2}
\begin{align}
\mathcal{A}_{F}  &  =M_{2}\left(  \mathbb{C}\right)  \oplus\mathbb{H},\qquad
n=2\\
\mathcal{A}_{F}  &  =M_{2}\left(  \mathbb{H}\right)  \oplus M_{4}\left(
\mathbb{C}\right)  ,\qquad n=4.
\end{align}
One, however, must remember that the maps $Y$ and $Y^{\prime}$ are functions
of the coordinates of the manifold $M$ and therefore the algebra associated
with this space must be
\begin{align}
\mathcal{A}  &  =C^{\infty}\left(  M,\mathcal{A}_{F}\right) \\
&  =C^{\infty}\left(  M\right)  \otimes\mathcal{A}_{F}.
\end{align}
To see this consider, for simplicity, the $n=2$ case with only the map $Y.$
The Clifford algebra $C_{-}=\mathbb{H}$ is spanned by the set $\left\{
1,\Gamma^{A}\right\}  ,$ $A=1,2,3,$ where $\left\{  \Gamma^{A},\Gamma
^{B}\right\}  =-2\delta^{AB}.$ We then consider functions which are made out
of words of the variable $Y$ formed with the use of constant elements of the
algebra \cite{Connes}
\[%
%TCIMACRO{\dsum \limits_{i=1}^{\infty}}%
%BeginExpansion
{\displaystyle\sum\limits_{i=1}^{\infty}}
%EndExpansion
a_{1}Ya_{2}Y\cdots a_{i}Y,\qquad a_{i}\in\mathbb{H},
\]
which will generate arbitrary functions over the manifold, which is the most
general form since $Y^{2}=1$. One can easily see that these combinations
generate all the spherical harmonics. This result could be easily generalized
by considering functions of the fields
\[
Z=\frac{1}{2}\left(  Y+1\right)  \left(  Y^{\prime}+1\right)  -1,\qquad
Y\in\mathbb{H},\quad Y^{\prime}\in M_{2}\left(  \mathbb{C}\right)  ,
\]
showing that the noncommutative algebra generated by the constant matrices and
the Feynman slash coordinates $Z$ is given by \cite{Connes}
\[
\mathcal{A}=C^{\infty}\left(  M_{2}\right)  \otimes\left(  \mathbb{H+}%
M_{2}\left(  \mathbb{C}\right)  \right)  .
\]

\section{Finite Noncommutative space}

Having explained the simple case $n=2$, for the remainder of this paper we
restrict ourselves to the physical case of $n=4.$ Here the algebra is given
by
\begin{equation}
\mathcal{A}=C^{\infty}\left(  M_{4}\right)  \otimes\left(  M_{2}%
(\mathbb{H})\mathbb{+}M_{4}\left(  \mathbb{C}\right)  \right)  .
\end{equation}
The associated Hilbert space is
\begin{equation}
\mathcal{H}=L^{2}\left(  M_{4},S\right)  \otimes\mathcal{H}_{F}.
\end{equation}
The Dirac operator mixes the finite space and the continuous manifold
non-trivially%
\begin{equation}
D=D_{M}\otimes1+\gamma_{5}\otimes D_{F},
\end{equation}
where $D_{F\text{ }}$ is a self adjoint operator in the finite space. The
chirality operator is
\begin{equation}
\gamma=\gamma_{5}\otimes\gamma_{F},
\end{equation}
and the anti-unitary operator $J$ is given by
\begin{equation}
J=J_{M}\gamma_{5}\otimes J_{F},
\end{equation}
where $J_{M}$ is the charge-conjugation operator $C$ on $M$ and $J_{F}$ the
anti-unitary operator for the finite space. Thus an element $\Psi
\in\mathcal{H}$ is of the form $\Psi=\left(
\begin{array}
[c]{c}%
\psi_{A}\\
\psi_{A^{\prime}}%
\end{array}
\right)  $ where $\psi_{A}$ is a $16$ component $L^{2}\left(  M,S\right)  $
spinor in the fundamental representation of $\mathcal{A}_{F}$ of the form
$\psi_{A}=\psi_{\alpha I}$ where $\alpha=1,\cdots,4$ with respect to
$M_{2}\left(  \mathbb{H}\right)  $ and $I=1,\cdots,4$ with respect to
$M_{4}\left(  \mathbb{C}\right)  $ and where $\psi_{A^{\prime}}=C\psi
_{A}^{\ast}$ is the charge conjugate spinor to $\psi_{A}$ \cite{AC}. The
chirality operator $\gamma$ must commute with elements of $\mathcal{A}$ which
implies that $\gamma_{F}$ must commute with elements in $\mathcal{A}_{F}.$
Commutativity of the chirality operator $\gamma_{F}$ with the algebra
$\mathcal{A}_{F}$ and that this $\mathbb{Z}/2$ grading acts non-trivially
reduces the algebra $M_{2}\left(  \mathbb{H}\right)  $ to $\mathbb{H}%
_{R}\oplus\mathbb{H}_{L}$ \cite{CC2}. Thus the $\gamma_{F}$ is identified with
$\gamma_{F}=\Gamma^{5}=\Gamma^{1}\Gamma^{2}\Gamma^{3}\Gamma^{4}$ and the
finite space algebra reduces to
\begin{equation}
\mathcal{A}_{F}=\mathbb{H}_{R}\oplus\mathbb{H}_{L}\oplus M_{4}\left(
\mathbb{C}\right)  .
\end{equation}
This can be easily seen by noting that an element of $M_{2}\left(
\mathbb{H}\right)  $ takes the form $\left(
\begin{array}
[c]{cc}%
q_{1} & q_{2}\\
q_{3} & q_{4}%
\end{array}
\right)  $ where each $q_{i},$ $i=1,\cdots,4,$ is a $2\times2$ matrix
representing a quaternion. Taking the representation of $\Gamma^{5}=\left(
\begin{array}
[c]{cc}%
1_{2} & 0\\
0 & -1_{2}%
\end{array}
\right)  $ to commute with $M_{2}\left(  \mathbb{H}\right)  $ implies that
$q_{2}=0=q_{3},$ thus reducing the algebra to $\mathbb{H}_{R}\oplus
\mathbb{H}_{L}.$ Therefore the index $\alpha=1,\cdots,4$ splits into two
parts, $\overset{.}{a}=\overset{.}{1},\overset{.}{2}$ which is a doublet under
$\mathbb{H}_{R}$ and $a=1,2$ which is a doublet under $\mathbb{H}_{L}$. The
spinor $\Psi$ further satisfies the chirality condition $\gamma\Psi=\Psi$
which implies that the spinors $\psi_{\overset{.}{a}I}$ are in the $\left(
2_{R},1_{L},4\right)  $ with respect to the algebra $\mathbb{H}_{R}%
\mathbb{\oplus H}_{L}\oplus M_{4}\left(  \mathbb{C}\right)  $ while $\psi
_{aI}$ are in the $\left(  1_{R},2_{L},4\right)  $ representation\footnote{Due
to a typographical error in the abstract of \cite{CCS2} the fermionic
representation was listed incorrectly as $\left(  2_{R},2_{L},4\right)  $
while in the body of the paper the coorect representation appears.}. The
finite space Dirac operator $D_{F}$ is then a $32\times32$ Hermitian matrix
acting on the $32$ component spinors $\Psi.$ In addition we take three copies
of each spinor to account for the three families, but will omit writing an
index for the families. At present we have no explanation for why the number
of generations should be three. The Dirac operator for the finite space is
then a $96\times96$ Hermitian matrix. The Dirac action is then given by
\cite{saddam}
\begin{equation}
\left(  J\Psi,D\Psi\right)  .
\end{equation}
We note that we are considering compact spaces with Euclidean signature and
thus the condition $J\Psi=\Psi$ could not be imposed. It could, however, be
imposed if the four dimensional space is Lorentzian \cite{Barrett}.The reason
is that the $KO$ dimension of the finite space is $6$ because the operators
$D_{F},$ $\gamma_{F}$ and $J_{F}$ satisfy%
\begin{equation}
J_{F}^{2}=1,\qquad J_{F}D_{F}=D_{F}J_{F},\qquad J_{F}\gamma_{F}=-\gamma
_{F}J_{F}.
\end{equation}
The operators $D_{M},$ $\gamma_{M}=\gamma_{5},$ and $J_{M}=C$ for a compact
manifold of dimension $4$ satisfy
\begin{equation}
J_{M}^{2}=-1,\qquad J_{M}D_{M}=D_{M}J_{M},\qquad J_{M}\gamma_{5}=\gamma
_{5}J_{M}. \label{Euclidean}%
\end{equation}
Thus the $KO$ dimension of the full noncommutative space $\left(
\mathcal{A},\mathcal{H},D\right)  $ with the decorations $J$ and $\gamma$
included is $10$ and satisfies
\begin{equation}
J^{2}=-1,\qquad JD=DJ,\qquad J\gamma=-\gamma J.
\end{equation}
We have shown in \cite{saddam} that the path integral of the Dirac action,
thanks to the relations $J^{2}=-1$ and $J\gamma=-\gamma J,$ \ \ \ \ yields a
Pfaffian of the operator $D$ instead of its determinant and thus eliminates
half the degrees of freedom of $\Psi$ and have the same effect as imposing the
condition $J\Psi=\Psi.$

We have also seen that the operator $J$ sends the algebra $\mathcal{A}$ to its
commutant, and thus the full algebra acting on the Hilbert space $\mathcal{H}$
is $\mathcal{A\otimes A}^{o}.$ Under automorphisms of the algebra
\begin{equation}
\Psi\rightarrow U\Psi,
\end{equation}
where $U=u\widehat{u}$ with $u\in\mathcal{A},$ $\widehat{u}\in\mathcal{A}^{o}$
with $\left[  u,\widehat{u}\right]  =0$, it is clear that Dirac action is not
invariant. This is similar to the situation in electrodynamics where the Dirac
action is not invariant under local phase transformations but the invariance
is easily restored by introducing the vector potential $A_{\mu}$ through the
transformation
\begin{equation}
\gamma^{\mu}\partial_{\mu}\rightarrow\gamma^{\mu}\left(  \partial_{\mu
}+ieA_{\mu}\right)  .
\end{equation}
In our case, the Dirac operator $D$ is replaced with
\begin{equation}
D_{A}=D+A,
\end{equation}
where the connection $A$ is given by \cite{CCS}
\begin{equation}
A=%
%TCIMACRO{\dsum }%
%BeginExpansion
{\displaystyle\sum}
%EndExpansion
a\widehat{a}\left[  D,b\widehat{b}\right]  .
\end{equation}
It can be shown that under automorphisms $U$ of the algebra we have
\begin{equation}
D_{A}\rightarrow UD_{A}U^{\ast}.
\end{equation}
The connection $A$ splits into three pieces%
\begin{equation}
A=A_{\left(  1\right)  }+JA_{(1)}J^{-1}+A_{\left(  2\right)  },
\end{equation}
where
\begin{align}
A_{\left(  1\right)  }  &  =%
%TCIMACRO{\dsum }%
%BeginExpansion
{\displaystyle\sum}
%EndExpansion
a\left[  D,b\right] \\
A_{\left(  2\right)  }  &  =%
%TCIMACRO{\dsum }%
%BeginExpansion
{\displaystyle\sum}
%EndExpansion
\widehat{a}\left[  A_{\left(  1\right)  },\widehat{b}\right]  ,
\end{align}
which satisfies $JA_{\left(  2\right)  }J^{-1}=A_{\left(  2\right)  }.$ At
this point we have to distinguish few possibilities.

\section{Pati-Salam Models}

In the first possibility we assume that the double commutator
\begin{equation}
\left[  a,\left[  D,\widehat{b}\right]  \right]  \neq0,
\end{equation}
which implies that $A_{\left(  2\right)  }\neq0.$ The fluctuations $A$ of the
inner automorphisms were computed in \cite{CCS2}. The calculation is
straightforward and could be easily done using symbolic manipulation programs
such as Mathematica or Maple. We shall content ourselves in this paper by
collecting some of the important results. Starting with $a\in M_{4}\left(
\mathbb{C}\right)  \oplus M_{4}\left(  \mathbb{C}\right)  $ we write
\begin{equation}
a=\left(
\begin{array}
[c]{cc}%
X_{\alpha}^{\beta}\delta_{I}^{J} & 0\\
0 & \delta_{\alpha^{\prime}}^{\beta^{\prime}}Y_{I^{\prime}}^{J^{\prime}}%
\end{array}
\right)  ,
\end{equation}
where $X_{\alpha}^{\beta}\in\mathbb{H}_{R}\mathbb{\oplus H}_{L}$ and
$Y_{I}^{J}\in M_{4}\left(  \mathbb{C}\right)  .$ Thus we now have
\begin{equation}
X_{\alpha}^{\beta}=\left(
\begin{array}
[c]{cc}%
X_{\overset{.}{a}}^{\overset{.}{b}} & 0\\
0 & X_{a}^{b}%
\end{array}
\right)  ,\qquad X_{a}^{b}=\left(
\begin{array}
[c]{cc}%
X_{1}^{1} & X_{1}^{2}\\
-\overline{X}_{1}^{2} & \overline{X}_{1}^{1}%
\end{array}
\right)  \in\mathbb{H}_{L},
\end{equation}
and similarly for $X_{\overset{.}{a}}^{\overset{.}{b}}\in\mathbb{H}_{R}$. The
anti-linear isometry $J=C\gamma_{5}\otimes J_{F}$ \ is represented by
\begin{equation}
J_{F}=\left(
\begin{array}
[c]{cc}%
0 & \delta_{\alpha}^{\beta^{\prime}}\delta_{I}^{J^{\prime}}\\
\delta_{\alpha^{\prime}}^{\beta}\delta_{I^{\prime}}^{J} & 0
\end{array}
\right)  \times\text{\textrm{complex conjugation,}}%
\end{equation}
and satisfies $J_{F}^{2}=1$ which implies that $J^{2}=-1$. In this form
\begin{equation}
\widehat{a}=Ja^{\ast}J^{-1}=\left(
\begin{array}
[c]{cc}%
\delta_{\alpha}^{\beta}Y_{I}^{tJ} & 0\\
0 & X_{\alpha^{\prime}}^{t\beta^{\prime}}\delta_{^{I^{\prime}\prime}%
}^{J^{\prime}}%
\end{array}
\right)
\end{equation}
where the superscript $t$ denotes the transpose matrix. This clearly satisfies
the commutation relation
\begin{equation}
\left[  a,\widehat{b}\right]  =0,
\end{equation}
which is simply the statement that the right action and left action commute.
We shall now show that the relations that $D$ must satisfy greatly constrain
its form. The (finite) Dirac operator can be written in matrix form%
\begin{equation}
D_{F}=\left(
\begin{array}
[c]{cc}%
D_{A}^{B} & D_{A}^{B^{^{\prime}}}\\
D_{A^{^{\prime}}}^{B} & D_{A^{^{\prime}}}^{B^{^{\prime}}}%
\end{array}
\right)  , \label{eq:dirac}%
\end{equation}
and must satisfy the properties
\begin{equation}
\gamma_{F}D_{F}=-D_{F}\gamma_{F}\qquad J_{F}D_{F}=D_{F}J_{F},
\end{equation}
where $J_{F}^{2}=1$. We also adopt the notation $D_{A^{\prime}}^{\quad
B}=D^{AB}.$

A\ matrix realization of $\gamma_{F}$ and $J_{F}$ is given by
\begin{equation}
\gamma_{F}=\left(
\begin{array}
[c]{cc}%
G_{F} & 0\\
0 & -\overline{G}_{F}%
\end{array}
\right)  ,\qquad G_{F}=\left(
\begin{array}
[c]{cc}%
1_{2} & 0\\
0 & -1_{2}%
\end{array}
\right)  ,\qquad J_{F}=\left(
\begin{array}
[c]{cc}%
0_{4} & 1_{4}\\
1_{4} & 0_{4}%
\end{array}
\right)  \circ\mathrm{cc.}%
\end{equation}
These relations, together with the hermiticity of $D$ imply the relations
\begin{equation}
\left(  D_{F}\right)  _{A^{^{\prime}}}^{B^{^{\prime}}}=\left(  \overline
{D}_{F}\right)  _{A}^{B}\,\qquad\left(  D_{F}\right)  _{A^{^{\prime}}}%
^{B}=\left(  \overline{D}_{F}\right)  _{B}^{A^{\prime}},
\end{equation}
with the bar denoting complex conjugation$.$ The operator $D_{F}$ have the
following zero components \cite{AC}
\begin{align}
\left(  D_{F}\right)  _{aI}^{bJ}  &  =0=\left(  D_{F}\right)  _{\overset{.}{a}%
I}^{\overset{.}{b}J}\\
\left(  D_{F}\right)  _{aI}^{\overset{.}{b}^{\prime}J^{\prime}}  &  =0=\left(
D_{F}\right)  _{\overset{.}{a}I}^{b^{\prime}J\prime},
\end{align}
leaving the components $\left(  D_{F}\right)  _{aI}^{\overset{.}{b}J}$,
$\left(  D_{F}\right)  _{aI}^{b^{\prime}J^{\prime}}$ and $\left(
D_{F}\right)  _{\overset{.}{a}I}^{\overset{.}{b}^{\prime}J^{\prime}}$
arbitrary. These restrictions lead to important constraints on the structure
of the connection that appears in the inner fluctuations of the Dirac operator.

We have shown, using elementary algebra, that the components of the connection
$A$ which is tensored with the Clifford gamma matrices $\gamma^{\mu}$ are the
gauge fields of the Pati-Salam model with the symmetry of $SU\left(  2\right)
_{R}\times SU\left(  2\right)  _{L}\times SU\left(  4\right)  .$ On the other
hand, the non-vanishing components of the connection which is tensored with
the gamma matrix $\gamma_{5}$ are given by
\begin{equation}
\left(  A\right)  _{aI}^{\overset{.}{b}J}\equiv\gamma_{5}\left(
\Sigma\right)  _{aI}^{\overset{.}{b}J},\qquad\left(  A\right)  _{aI}%
^{b^{\prime}J^{\prime}}=\gamma_{5}H_{aIbJ},\qquad\left(  A\right)
_{\overset{.}{a}I}^{\overset{.}{b}^{\prime}J^{\prime}}\equiv\gamma
_{5}H_{\overset{.}{a}I\overset{.}{b}J},
\end{equation}
where $H_{aIbJ}=H_{bJaI}$ and $H_{\overset{.}{a}I\overset{.}{b}J}%
=H_{\overset{.}{b}J\overset{.}{a}I}$ , which is the most general Higgs
structure possible. These correspond to the representations with respect to
$SU\left(  2\right)  _{R}\times SU\left(  2\right)  _{L}\times SU\left(
4\right)  $ \cite{CCS2}%
\begin{align}
\Sigma_{aI}^{\overset{.}{b}J}  &  =\left(  \overline{2}_{R},2_{L},1\right)
+\left(  \overline{2}_{R},2_{L},15\right) \\
H_{aIbJ}  &  =\left(  1_{R},1_{L},6\right)  +\left(  1_{R},3_{L},10\right) \\
H_{\overset{.}{a}I\overset{.}{b}J}  &  =\left(  1_{R},1_{L},6\right)  +\left(
3_{R},1_{L},10\right)  .
\end{align}
We note, however, that the inner fluctuations form a semi-group and if a
component $\left(  D_{F}\right)  _{aI}^{\overset{.}{b}J}$ or $\left(
D_{F}\right)  _{aI}^{b^{\prime}J^{\prime}}$ or $\left(  D_{F}\right)
_{\overset{.}{a}I}^{\overset{.}{b}^{\prime}J^{\prime}}$ vanish, then the
corresponding $A$ field will also vanish. We distinguish three cases: 1)
Left-right symmetric Pati--Salam model with fundamental Higgs fields
$\Sigma_{aI}^{\overset{.}{b}J},$ $H_{aIbJ}$ and $H_{\overset{.}{a}%
I\overset{.}{b}J}.$ In this model the field $H_{aIbJ}$ should have a zero vev.
2) A Pati-Salam model where the Higgs field $H_{aIbJ}$ that couples to the
left sector is set to zero (and then remain zero under fluctuations) which is
desirable because there is no symmetry between the left and right sectors at
low energies. 3) The initial values for $\left(  D_{F}\right)  _{aI}%
^{\overset{.}{b}J}$ , $\left(  D_{F}\right)  _{aI}^{b^{\prime}J^{\prime}}$ and
$\left(  D_{F}\right)  _{\overset{.}{a}I}^{\overset{.}{b}^{\prime}J^{\prime}}$
before fluctuations are given by those that are determined for the Standard
Model, where order one condition is satisfied for the subalgebra, then the
Higgs fields $\Sigma_{aI}^{\overset{.}{b}J},$ $H_{aIbJ}$ and
$H_{\overset{.}{a}I\overset{.}{b}J}$ will become dependent fields and
expressible in terms of more fundamental fields (as will be shown in the next
section). \ \ \ \ \ \ \ \ \ \ \ \ \ \ \ \ \ \ \ \ \ \ \ \ \ \ \ \ \ \ \ \ \ \ \ \ \ \ \ \ \ \ \ \ \ \ \ \ \ \ \ \ \ \ \ \ \ \ \ \ \ \ 

In matrix form the operator $D_{F}$ has the sub-matrices \cite{AC}
\begin{equation}
\left(  D_{F}\right)  _{\alpha I}^{\quad\beta J}=\left(
\begin{array}
[c]{cc}%
0 & D_{aI}^{\overset{.}{b}J}\\
D_{\overset{.}{a}I}^{bJ} & 0
\end{array}
\right)  ,\qquad D_{aI}^{\overset{.}{b}J}=\left(  D_{\overset{.}{a}J}%
^{bI}\right)  ^{\ast}.
\end{equation}
Then the components of the Dirac operator tensored with $\gamma^{\mu},$
including inner fluctuations, is given by \cite{CCS2}%
\begin{align}
\left(  D_{A}\right)  _{\overset{.}{a}I}^{\overset{.}{b}J}  &  =\gamma^{\mu
}\left(  D_{\mu}\delta_{\overset{.}{a}}^{\overset{.}{b}}\delta_{I}^{J}%
-\frac{i}{2}g_{R}W_{\mu R}^{\alpha}\left(  \sigma^{\alpha}\right)
_{\overset{.}{a}}^{\overset{.}{b}}\delta_{I}^{J}-\delta_{\overset{.}{a}%
}^{\overset{.}{b}}\left(  \frac{i}{2}gV_{\mu}^{m}\left(  \lambda^{m}\right)
_{I}^{^{J}}+\frac{i}{2}gV_{\mu}\delta_{I}^{J}\right)  \right) \\
\left(  D_{A}\right)  _{aI}^{bJ}  &  =\gamma^{\mu}\left(  D_{\mu}\delta
_{a}^{b}\delta_{I}^{J}-\frac{i}{2}g_{L}W_{\mu L}^{\alpha}\left(
\sigma^{\alpha}\right)  _{a}^{b}\delta_{I}^{J}-\delta_{a}^{b}\left(  \frac
{i}{2}gV_{\mu}^{m}\left(  \lambda^{m}\right)  _{I}^{^{J}}+\frac{i}{2}gV_{\mu
}\delta_{I}^{J}\right)  \right)  ,
\end{align}
where the fifteen $4\times4$ matrices $\left(  \lambda^{m}\right)  _{I}^{^{J}%
}$ are traceless and generate the group $SU\left(  4\right)  $ and $W_{\mu
R}^{\alpha},$ $W_{\mu L}^{\alpha},$ $V_{\mu}^{m}$ are the gauge fields of
$SU\left(  2\right)  _{R}$, $SU\left(  2\right)  _{L}$, and $SU\left(
4\right)  .$ The requirement that $A$ is unimodular implies that
\begin{equation}
\mathrm{Tr}\left(  A\right)  =0,
\end{equation}
which gives the condition
\begin{equation}
V_{\mu}=0.
\end{equation}
This shows that the resulting gauge group is $SU\left(  2\right)  _{R}\times
SU\left(  2\right)  _{L}\times SU\left(  4\right)  $, which is the Pati-Salam
gauge symmetry. In addition we have for the components of the Dirac operator
tensored with $\gamma_{5},$
\begin{align}
\left(  D_{A}\right)  _{\overset{.}{a}I}^{bJ}  &  =\gamma_{5}\Sigma
_{\overset{.}{a}I}^{bJ}\\
\left(  D_{A}\right)  _{\overset{.}{a}I}^{\overset{.}{b}^{\prime}J^{\prime}}
&  =\gamma_{5}H_{\overset{.}{a}I\overset{.}{b}J}\\
\left(  D_{A}\right)  _{aI}^{b^{\prime}J^{\prime}}  &  =\gamma_{5}H_{aIbJ},
\end{align}
where $\Sigma_{\overset{.}{a}I}^{bJ}$ is in the $\left(  2_{R},2_{L}%
,1+15\right)  $ representation, $H_{\overset{.}{a}I\overset{.}{b}%
J}=H_{\overset{.}{b}J\overset{.}{a}I}$ is in the $\left(  3_{R},1_{L}%
,10\right)  +\left(  1_{R},1_{L},6\right)  $ representation and $H_{aIbJ}$ is
in the $\left(  1_{R},1_{L},6\right)  +\left(  1_{R},3_{L},10\right)  $ with
respect to $SU\left(  2\right)  _{R}\times SU\left(  2\right)  _{L}\times
SU\left(  4\right)  .$ To conclude, there are only three Pati-Salam models
with fixed Higgs structure, where the first one is the most general case, and
the other two are special cases of the first one.

\section{The Standard Model}

We now consider the situation when the order one condition is satisfied
\begin{equation}
\left[  a,\left[  D,\widehat{b}\right]  \right]  =0,
\end{equation}
and the center of the algebra $Z\left(  \mathcal{A}\right)  $ is non-trivial
in such a way that the space is connected. Physically, this means that there
is a mixing term between the fermions and their conjugates. The Dirac operator
connects the spinors $\psi_{A}$ and their conjugates $\psi_{A^{\prime}}$ so
that
\begin{equation}
\left[  D,Z\left(  \mathcal{A}\right)  \right]  \neq0.
\end{equation}
In physical terms this would allow a Majorana mass term for the fermions. It
was shown in \cite{CC2} that the unique solution to this equation constrains
the algebra $\mathcal{A}_{F}=\mathbb{H}_{R}\oplus\mathbb{H}_{L}\oplus
M_{4}\left(  \mathbb{C}\right)  $ to be restricted to a subalgebra
\begin{equation}
\mathbb{C}\oplus\mathbb{H}_{L}\oplus M_{3}\left(  \mathbb{C}\right)  ,
\end{equation}
so that an element of $\mathcal{A}$ takes the form \cite{AC}
\begin{equation}
a=\left(
\begin{array}
[c]{ccccc}%
X\otimes1_{4} &  &  &  & \\
& \overline{X}\otimes1_{4} &  &  & \\
&  & q\otimes1_{4} &  & \\
&  &  & 1_{4}\otimes X & \\
&  &  &  & 1_{4}\otimes m
\end{array}
\right)  ,\qquad.
\end{equation}
where $X\in\mathbb{C},\quad q\in\mathbb{H},\quad$ $m\in M_{3}\left(
\mathbb{C}\right)  $ and the operator $D_{F}$ have a singlet non-zero entry in
the mixing term $\left(  D_{F}\right)  _{A}^{A^{\prime}}$%
\begin{align}
\left(  D_{F}\right)  _{\alpha I}^{\beta J}  &  =\left(  \delta_{\alpha}%
^{1}\delta_{\overset{.}{1}}^{\beta}k^{\ast\nu}+\delta_{\alpha}^{\overset{.}{1}%
}\delta_{1}^{\beta}k^{\nu}+\delta_{\alpha}^{2}\delta_{\overset{.}{2}}^{\beta
}k^{\ast e}+\delta_{\alpha}^{\overset{.}{2}}\delta_{2}^{\beta}k^{e}\right)
\delta_{I}^{1}\delta_{1}^{J}\label{DSM1}\\
&  +\left(  \delta_{\alpha}^{1}\delta_{\overset{.}{1}}^{\beta}k^{\ast
u}+\delta_{\alpha}^{\overset{.}{1}}\delta_{1}^{\beta}k^{u}+\delta_{\alpha}%
^{2}\delta_{\overset{.}{2}}^{\beta}k^{\ast d}+\delta_{\alpha}^{\overset{.}{2}%
}\delta_{2}^{\beta}k^{d}\right)  \delta_{I}^{i}\delta_{j}^{J}\delta_{i}%
^{j}\nonumber\\
\left(  D_{F}\right)  _{\alpha I}^{\beta^{\prime}K^{\prime}}  &
=\delta_{\alpha}^{\overset{.}{1}}\delta_{\overset{.}{1}^{\prime}}%
^{\beta^{\prime}}\delta_{I}^{1}\delta_{1^{\prime}}^{K^{\prime}}k^{\ast\nu_{R}%
}\sigma, \label{DSM2}%
\end{align}
where $k^{\nu},k^{e},$ $k^{u},$ $k^{d}$ and $k^{\nu_{R}}$ are $3\times3$
Yukawa couplings in generation space. The field $\sigma$ is a singlet (which
could be complex) whose vev is responsible for the right-handed neutrino
Majorana mass. The operator $D$ must be replaced with the operator
\begin{equation}
D_{A}=D+A+JAJ^{-1},
\end{equation}
and
\begin{equation}
A_{\left(  2\right)  }=0,
\end{equation}
which greatly simplifies the Higgs structure. The various components of the
Dirac operator are exactly those of the Standard Model, in addition to the
Higgs fields which are the components of the connection $A$ along discrete
directions%
\begin{align*}
\left(  D\right)  _{\overset{.}{1}1}^{\overset{.}{1}1}  &  =\gamma^{\mu
}\otimes D_{\mu}\otimes1_{3},\quad D_{\mu}=\partial_{\mu}+\frac{1}{4}%
\omega_{\mu}^{cd}\left(  e\right)  \gamma_{cd},\quad1_{3}=\text{generations}\\
\left(  D\right)  _{\overset{.}{1}1}^{a1}  &  =\gamma_{5}\otimes k^{\ast\nu
}\otimes\epsilon^{ab}H_{b}\qquad k^{\nu}=3\times3\text{ neutrino mixing
matrix}\\
\left(  D\right)  _{\overset{.}{2}1}^{\overset{.}{2}1}  &  =\gamma^{\mu
}\otimes\left(  D_{\mu}+ig_{1}B_{\mu}\right)  \otimes1_{3}\\
\left(  D\right)  _{\overset{.}{2}1}^{a1}  &  =\gamma_{5}\otimes k^{\ast
e}\otimes\overline{H}^{a}\\
\left(  D\right)  _{a1}^{\overset{.}{1}1}  &  =\gamma_{5}\otimes k^{\nu
}\otimes\epsilon_{ab}\overline{H}^{b}\\
\left(  D\right)  _{a1}^{\overset{.}{2}1}  &  =\gamma_{5}\otimes k^{e}\otimes
H_{a}\\
\left(  D\right)  _{a1}^{b1}  &  =\gamma^{\mu}\otimes\left(  \left(  D_{\mu
}+\frac{i}{2}g_{1}B_{\mu}\right)  \delta_{a}^{b}-\frac{i}{2}g_{2}W_{\mu
}^{\alpha}\left(  \sigma^{\alpha}\right)  _{a}^{b}\right)  \otimes1_{3},\text{
\qquad}\sigma^{\alpha}=\text{Pauli}\\
\left(  D\right)  _{\overset{.}{1}i}^{\overset{.}{1}j}  &  =\gamma^{\mu
}\otimes\left(  \left(  D_{\mu}-\frac{2i}{3}g_{1}B_{\mu}\right)  \delta
_{i}^{j}-\frac{i}{2}g_{3}V_{\mu}^{m}\left(  \lambda^{m}\right)  _{i}%
^{j}\right)  \otimes1_{3},\qquad\lambda^{i}=\text{Gell-Mann}\\
\left(  D\right)  _{\overset{.}{1}i}^{aj}  &  =\gamma_{5}\otimes k^{\ast
u}\otimes\epsilon^{ab}H_{b}\delta_{i}^{j}\\
\left(  D\right)  _{\overset{.}{2}i}^{\overset{.}{2}j}  &  =\gamma^{\mu
}\otimes\left(  \left(  D_{\mu}+\frac{i}{3}g_{1}B_{\mu}\right)  \delta_{i}%
^{j}-\frac{i}{2}g_{3}V_{\mu}^{m}\left(  \lambda^{m}\right)  _{i}^{j}\right)
\otimes1_{3}\\
\left(  D\right)  _{\overset{.}{2}i}^{aj}  &  =\gamma_{5}\otimes k^{\ast
d}\otimes\overline{H}^{a}\delta_{i}^{j}\\
\left(  D\right)  _{ai}^{bj}  &  =\gamma^{\mu}\otimes\left(  \left(  D_{\mu
}-\frac{i}{6}g_{1}B_{\mu}\right)  \delta_{a}^{b}\delta_{i}^{j}-\frac{i}%
{2}g_{2}W_{\mu}^{\alpha}\left(  \sigma^{\alpha}\right)  _{a}^{b}\delta_{i}%
^{j}-\frac{i}{2}g_{3}V_{\mu}^{m}\left(  \lambda^{m}\right)  _{i}^{j}\delta
_{a}^{b}\right)  \otimes1_{3}\\
\left(  D\right)  _{ai}^{\overset{.}{1}j}  &  =\gamma_{5}\otimes k^{u}%
\otimes\epsilon_{ab}\overline{H}^{b}\delta_{i}^{j}\\
\left(  D\right)  _{ai}^{\overset{.}{2}j}  &  =\gamma_{5}\otimes k^{d}\otimes
H_{a}\delta_{i}^{j}\\
\left(  D\right)  _{\overset{.}{1}1}^{\overset{.}{1^{\prime}}1^{\prime}}  &
=\gamma_{5}\otimes k^{\ast\nu_{R}}\sigma\qquad\text{generate scale }%
M_{R}\text{ by }\sigma\rightarrow M_{R}\\
\left(  D\right)  _{\overset{.}{1^{\prime}}1^{\prime}}^{\overset{.}{1}1}  &
=\gamma_{5}\otimes k^{\nu_{R}}\sigma\\
D_{A^{\prime}}^{B^{\prime}}  &  =\overline{D}_{A}^{B},\qquad D_{A^{\prime}%
}^{B}=\overline{D}_{A}^{B^{\prime}},\qquad D_{A}^{B^{\prime}}=\overline
{D}_{A^{\prime}}^{B}%
\end{align*}
where in this notation the fermions are enumerated as
\begin{align}
\psi_{\overset{.}{1}1}  &  =\nu_{R}\\
\psi_{\overset{.}{2}1}  &  =e_{R}\\
\psi_{a1}  &  =l_{a}=\left(
\begin{array}
[c]{c}%
\nu_{L}\\
e_{L}%
\end{array}
\right) \\
\psi_{\overset{.}{1}i}  &  =u_{iR}\\
\psi_{\overset{.}{2}i}  &  =d_{iR}\\
\psi_{ai}  &  =q_{ia}=\left(
\begin{array}
[c]{c}%
u_{iL}\\
d_{iL}%
\end{array}
\right)  .
\end{align}
It is clear that the associated gauge group is $U\left(  1\right)  \times
SU\left(  2\right)  \times SU\left(  3\right)  $ and that there is only one
Higgs doublet $H.$ We note the presence of the singlet field $\sigma$ which is
the field whose vev will give a Majorana mass to the right-handed neutrinos.
This field plays an essential role in stabilizing the Higgs coupling so that
it does not turn negative at very high energies \cite{Reselience}. We \ note
in passing that the number of generations is inserted by hand in the Dirac
operator of the finite space, and at present we do not have any geometrical
explanation to single out three generations.

\section{A special Pati-Salam model}

We have shown that inner fluctuations resulting from the action on operators
in Hilbert space form a semi-group Pert$\left(  \mathcal{A}\right)  .$ There
exists configurations for which the inverse transformation to the perturbation
does not exist. One such Dirac operator $D_{A}$ corresponds to the case where
the initial operator $D$ is taken to be the one deduced for the Standard Model
as given in (\ref{DSM1}) and (\ref{DSM2}), but not restricting its action to
the subalgebra $\mathbb{C}\oplus\mathbb{H}_{L}\oplus M_{3}\left(
\mathbb{C}\right)  $ but to the full algebra $\mathbb{H}_{R}\oplus
\mathbb{H}_{L}\oplus M_{4}\left(  \mathbb{C}\right)  .$ In this case one finds
out that the resultant vector fields are the same as in the case of Pati-Salam
models, but where the Higgs fields $\Sigma_{\overset{.}{a}I}^{bJ}$ and
$H_{\overset{.}{a}I\overset{.}{b}J}$ become composite fields determined in
function of fundamental Higgs fields while $H_{aIbJ}$ vanishes. These are
given by \cite{CCS2}
\begin{align}
\Sigma_{\overset{.}{a}I}^{bJ}  &  =\left(  \left(  k^{\nu}\phi_{\overset{.}{a}%
}^{b}+k^{e}\widetilde{\phi}_{\overset{.}{a}}^{b}\right)  \Sigma_{I}%
^{J}+\left(  k^{u}\phi_{\overset{.}{a}}^{b}+k^{d}\widetilde{\phi
}_{\overset{.}{a}}^{b}\right)  \left(  \delta_{I}^{J}-\Sigma_{I}^{J}\right)
\right) \\
H_{\overset{.}{a}I\overset{.}{b}J}  &  =k^{\ast\nu_{R}}\Delta_{\overset{.}{a}%
J}\Delta_{\overset{.}{b}I}\\
H_{aIbJ}  &  =0,
\end{align}
where the Higgs field $\phi_{\overset{.}{a}}^{b}$ is in the $\left(
2_{R},\overline{2}_{L},1\right)  $ of the product gauge group $SU\left(
2\right)  _{R}\times SU\left(  2\right)  _{L}\times SU\left(  4\right)  $,
$\widetilde{\phi}_{\overset{.}{a}}^{b}=\tau_{2}\overline{\phi}_{\overset{.}{a}%
}^{b}\tau_{2}$ and $\Delta_{\overset{.}{a}J}$ is in the $\left(  2_{R,}%
,1_{L},4\right)  $ representation while $\Sigma_{I}^{J}$ is in the $\left(
1_{R},1_{L},1+15\right)  $ representation. The fact that one gets a simpler
Higgs representations in this case makes it more attractive. It is certainly
an interesting question to determine all Dirac operators which lead to
singular transformations where the resultant Higgs fields are composites of
more fundamental ones. The scalar potential which contains quartic
interactions in the bosonic fields, which because of compositness, are of
order $8.$ All terms of orders higher than four will be suppressed by the
cut-off scale and could be truncated. Similarly the coupling of such terms to
the fermionic fields will be suppressed by the cut-off scale. To conclude this
section, it is remarkable that starting with the simple quantization condition
which represents the Chern-character of the noncommutative space and is a
special case of the orientability condition, fixes uniquely the structure of
space-time as well as the matter content in the form of a very specific
Pati-Salam unification model, or three of its truncations, including the
Standard Model. This enables us to track gravitational and matter
interactions, starting from the Planck scale where the starting point is few
spheres of Planck size, and ending up with the present scale. This compelling
picture could represent a valid framework for the realization of Hilbert's
program for axiomatization of physics.

\section{Spectral Action}

The coordinates $Y^{A}\left(  x\right)  $ are topological fields, and apart
from being coordinates of a sphere and satisfying the volume quantization
condition, are not constrained. They do play a role serving as coordinates
conjugate to the momentum represented by the Dirac operator. In particular,
since now $D$ and $Y$ play the role of momenta and coordinates, it is natural
to consider the spectral action to be of the form \cite{CC2}%
\[
\mathrm{Tr}f\left(  D_{A},Y\right)  ,
\]
which, because $Y^{2}=1,$ implies the dependence on terms of the form $\left[
D,Y\right]  .$ The lowest order contribution of such terms come from $\left[
D,Y\right]  ^{2}$ which corresponds to adding the following term to the
action
\begin{equation}
\frac{1}{2}%
%TCIMACRO{\dint \limits_{M}}%
%BeginExpansion
{\displaystyle\int\limits_{M}}
%EndExpansion
d^{4}x\sqrt{g}g^{\mu\nu}\partial_{\mu}Y^{A}\partial_{\nu}Y^{A}.
\end{equation}
It is also clear that in the case of the two sided quantization with the field
$Z$ the contribution of the term $\left[  D,Z\right]  ^{2}$ gives the sum of
two contributions without interference terms
\[
\frac{1}{2}%
%TCIMACRO{\dint \limits_{M}}%
%BeginExpansion
{\displaystyle\int\limits_{M}}
%EndExpansion
d^{4}x\sqrt{g}g^{\mu\nu}\left(  \partial_{\mu}Y^{A}\partial_{\nu}%
Y^{A}+\partial_{\mu}Y^{^{\prime}A}\partial_{\nu}Y^{^{\prime}A}\right)  .
\]
We have shown that the spectral action for the part dependent on $D_{A}$ gives
the bosonic action for all dynamical fields appearing in the connection $A.$
In particular, in the case of the Standard Model the bosonic action for the
part independent of the fields $Y^{A}$ and $Y^{\prime A}$ is given by
\cite{AC2} \cite{saddam} \cite{AC} \cite{CC}
\begin{equation}
S=2f_{4}\Lambda^{4}a_{0}+2f_{2}\Lambda^{2}a_{2}+f_{0}a_{4}+\cdots
\end{equation}
and
\begin{align}
S_{\mathrm{b}}  &  =\frac{48}{\pi^{2}}f_{4}\Lambda^{4}%
%TCIMACRO{\dint }%
%BeginExpansion
{\displaystyle\int}
%EndExpansion
d^{4}x\sqrt{g}\\
&  -\frac{4}{\pi^{2}}f_{2}\Lambda^{2}%
%TCIMACRO{\dint }%
%BeginExpansion
{\displaystyle\int}
%EndExpansion
d^{4}x\sqrt{g}\left(  R+\frac{1}{2}a\overline{H}H+\frac{1}{4}c\sigma
^{2}\right) \nonumber\\
&  +\frac{1}{2\pi^{2}}f_{0}%
%TCIMACRO{\dint }%
%BeginExpansion
{\displaystyle\int}
%EndExpansion
d^{4}x\sqrt{g}\left[  \frac{1}{30}\left(  -18C_{\mu\nu\rho\sigma}%
^{2}+11R^{\ast}R^{\ast}\right)  +\frac{5}{3}g_{1}^{2}B_{\mu\nu}^{2}+g_{2}%
^{2}\left(  W_{\mu\nu}^{\alpha}\right)  ^{2}+g_{3}^{2}\left(  V_{\mu\nu}%
^{m}\right)  ^{2}\right. \nonumber\\
&  \qquad\left.  +\frac{1}{6}aR\overline{H}H+b\left(  \overline{H}H\right)
^{2}+a\left\vert \nabla_{\mu}H_{a}\right\vert ^{2}+2e\overline{H}H\,\sigma
^{2}+\frac{1}{2}d\,\sigma^{4}+\frac{1}{12}cR\sigma^{2}+\frac{1}{2}c\left(
\partial_{\mu}\sigma\right)  ^{2}\right] \nonumber\\
&  +\cdots\nonumber
\end{align}
where $a,c,d,e$ are defined in terms of the Yukawa couplings, $f_{0}=f\left(
0\right)  $ and $f_{k}$ are the Mellin transforms of the function $f$%
\begin{equation}
f_{k}=%
%TCIMACRO{\dint \limits_{0}^{\infty}}%
%BeginExpansion
{\displaystyle\int\limits_{0}^{\infty}}
%EndExpansion
f\left(  v\right)  v^{k-1}dv,\qquad k>0.
\end{equation}
This action is calculated using heat kernel methods and was shown to contain
unification of gravity with gauge symmetries and Higgs field and the scalar
singlet. All couplings are related at unification scale. The zeroth order term
in the expansion gives the cosmological constant, the first order gives the
Einstein-Hilbert action and the scalar masses, and the second order gives the
Yang-Mills and scalar kinetic terms as well as the second order in curvature
terms. The presence of the singlet field $\sigma$ whose vev gives mass to the
right-handed neutrino plays an important role in stabilizing the Higgs
coupling which will not become negative at very high energies as well as being
consistent with a low Higgs mass of $126$ Gev \cite{Reselience}. The form of
the gauge and Higgs kinetic terms and potential implies unification of the
gauge couplings and the Higgs coupling. In addition there is a relation
between the fermion masses and the gauge field masses. A study of the RGE
showed that these relations are consistent with present experimental data and
predicts the top quark mass to be around $170$ Gev. However, gauge coupling
unification is off by $4\%$ indicating that the Standard Model is an excellent
approximation to a Pati-Salam model listed above. We have shown \cite{Walter}
that gauge coupling unification is indeed possible for Pati-Salam models at a
unification scale of the order of $10^{16}$ Gev.

It is also worthwhile to summarize the fermionic action
\begin{align}
S_{\mathrm{f}}  &  =%
%TCIMACRO{\dint }%
%BeginExpansion
{\displaystyle\int}
%EndExpansion
d^{4}x\sqrt{g}\left(  \nu_{R}^{\ast}\gamma^{\mu}D_{\mu}\nu_{R}\right. \\
+  &  e_{R}^{\ast}\gamma^{\mu}\left(  D_{\mu}+ig_{1}B_{\mu}\right)
e_{R}\nonumber\\
+  &  l_{L}^{a\ast}\gamma^{\mu}\left(  \left(  D_{\mu}+\frac{i}{2}g_{1}B_{\mu
}\right)  \delta_{a}^{b}-\frac{i}{2}g_{2}W_{\mu}^{\alpha}\left(
\sigma^{\alpha}\right)  _{a}^{b}\right)  l_{_{b}L}\nonumber\\
+  &  u_{R\ }^{i\ast}\gamma^{\mu}\left(  \left(  D_{\mu}-\frac{2i}{3}%
g_{1}B_{\mu}\right)  \delta_{i}^{j}-\frac{i}{2}g_{3}V_{\mu}^{m}\left(
\lambda^{m}\right)  _{i}^{j}\right)  u_{jR}\nonumber\\
+  &  d_{R}^{i\ast}\gamma^{\mu}\left(  \left(  D_{\mu}+\frac{i}{3}g_{1}B_{\mu
}\right)  \delta_{i}^{j}-\frac{i}{2}g_{3}V_{\mu}^{m}\left(  \lambda
^{m}\right)  _{i}^{j}\right)  d_{jR}\nonumber\\
+  &  q_{L}^{ia\ast}\gamma^{\mu}\left(  \left(  D_{\mu}-\frac{i}{6}g_{1}%
B_{\mu}\right)  \delta_{a}^{b}\delta_{i}^{j}-\frac{i}{2}g_{2}W_{\mu}^{\alpha
}\left(  \sigma^{\alpha}\right)  _{a}^{b}\delta_{i}^{j}-\frac{i}{2}g_{3}%
V_{\mu}^{m}\left(  \lambda^{m}\right)  _{i}^{j}\delta_{a}^{b}\right)
q_{jbL}\nonumber\\
&  +\nu_{R}^{\ast}\gamma_{5}k^{\ast\nu}\epsilon^{ab}H_{b}l_{_{a}L}+e_{R}%
^{\ast}\gamma_{5}k^{\ast e}\overline{H}^{a}l_{_{a}L}\nonumber\\
&  \left.  +u_{R\ }^{i\ast}\gamma_{5}k^{\ast u}\epsilon^{ab}H_{b}\delta
_{i}^{j}q_{jaL}+d_{R}^{i\ast}\gamma_{5}k^{\ast d}\overline{H}^{a}\delta
_{i}^{j}q_{jaL}+\nu_{R}^{\ast}\gamma_{5}k^{\ast\nu_{R}}\sigma\left(  \nu
_{R}^{\ast}\right)  ^{c}+\mathrm{h.c}\right)  .\nonumber
\end{align}
Note that the singlet field $\sigma$ after getting a vev from the minima of
its potential, will give a Majorana mass to the right-handed neutrino and
implies that the left handed neutrino will have a small mass through a see-saw mechanism.

\section{Consequences of volume quantization}

Having established the importance of the volume quantization condition, which
in turn implies that the two sets of fields $Y$ and $Y^{\prime}$ mapping the
four dimensional manifold to four spheres must be taken into consideration
when studying the dynamical content of the resulting model. In particular, the
Einstein equations of motion will be modified. The volume constraint, imposed
through a Lagrange multiplier, will result in traceless Einstein equations,
with the trace part equated to the Lagrange multiplier. We will show that
Bianchi identities give rise to a cosmological constant as an integration
constant. We now study the implications of the presence of the fields $Y$ and
$Y^{\prime}$ on the structure of the model.

For simplicity and to avoid cluttering of fields and indices, in what follows
we shall consider only one set of fields $Y^{A}$ and not two sets $Y^{A}$ and
$Y^{^{\prime}A}$ as required by the reality condition. The effects on the
equations of motion will \ be minimal. Here we take $Y\in M_{2}\left(
\mathbb{H}\right)  $ a $2\times2$ matrix whose elements are quaternions. This
can be written as
\begin{equation}
Y=Y^{A}\Gamma_{A},\qquad A=1,\cdots,5,
\end{equation}
where $\Gamma_{A}$ are Hermitian gamma matrices satisfying $\left\{
\Gamma^{A},\Gamma^{B}\right\}  =2\delta^{AB}$ where Cliff$\left(
+,+,+,+,+\right)  =M_{2}\left(  \mathbb{H}\right)  \oplus M_{2}\left(
\mathbb{H}\right)  $ and we take one of the irreducible representations
$M_{2}\left(  \mathbb{H}\right)  .$ The condition $Y^{2}=1$ implies
\begin{equation}
Y^{A}Y^{A}=1,
\end{equation}
which defines coordinates on the four dimensional sphere $S^{4}.$ We can check
that%
\begin{equation}
\frac{1}{2^{2}(4!)}\left\langle Y\left[  D,Y\right]  \left[  D,Y\right]
\left[  D,Y\right]  \left[  D,Y\right]  \right\rangle =\gamma,
\end{equation}
implies the relation
\begin{equation}
\det\left(  e_{\mu}^{a}\right)  =\frac{1}{4!}\epsilon^{\mu\nu\kappa\lambda
}\epsilon_{ABCDE}Y^{A}\partial_{\mu}Y^{B}\partial_{\nu}Y^{C}\partial_{\kappa
}Y^{D}\partial_{\lambda}Y^{E},
\end{equation}
which fixes the volume density and whose integral quantizes the volume. This
last condition can be imposed through a Lagrange multiplier. To do this
consider the action
\begin{align}
I  &  =-\frac{1}{2\kappa^{2}}%
%TCIMACRO{\dint }%
%BeginExpansion
{\displaystyle\int}
%EndExpansion
d^{4}x\sqrt{g}R+\frac{1}{2}%
%TCIMACRO{\dint }%
%BeginExpansion
{\displaystyle\int}
%EndExpansion
d^{4}x\lambda\left(  \frac{1}{\kappa^{4}}\sqrt{g}-\frac{1}{4!}\epsilon^{\mu
\nu\kappa\lambda}\epsilon_{ABCDE}Y^{A}\partial_{\mu}Y^{B}\partial_{\nu}%
Y^{C}\partial_{\kappa}Y^{D}\partial_{\lambda}Y^{E}\right) \nonumber\\
&  +\frac{1}{2\kappa^{4}}%
%TCIMACRO{\dint }%
%BeginExpansion
{\displaystyle\int}
%EndExpansion
d^{4}x\sqrt{g}\lambda^{\prime}\left(  Y^{A}Y^{A}-1\right)  ,
\end{align}
where $\kappa^{2}=8\pi G$ which will be set to $1.$ Notice that the third term
is a four-form and represents the volume element of a unit four-sphere. It can
be written in terms of differential forms without any tensor indices%
\begin{align}
&  -\frac{1}{2(4!)}%
%TCIMACRO{\dint }%
%BeginExpansion
{\displaystyle\int}
%EndExpansion
\lambda\epsilon_{ABCDE}Y^{A}dY^{B}\wedge dY^{C}\wedge dY^{D}\wedge dY^{E}\\
&  =-\frac{1}{8(4!)}%
%TCIMACRO{\dint }%
%BeginExpansion
{\displaystyle\int}
%EndExpansion
\lambda\mathrm{Tr}\left(  YdY\wedge dY\wedge dY\wedge dY\right)  ,
\end{align}
and is independent of the variation of the metric. Varying the action with
respect to the metric, after imposing the two Lagrange multipliers constraints%
\begin{align}
Y^{A}Y^{A}  &  =1\\
\sqrt{g}  &  =\frac{1}{4!}\epsilon^{\mu\nu\kappa\lambda}\epsilon_{ABCDE}%
Y^{A}\partial_{\mu}Y^{B}\partial_{\nu}Y^{C}\partial_{\kappa}Y^{D}%
\partial_{\lambda}Y^{E}, \label{volume}%
\end{align}
gives%
\begin{equation}
G_{\mu\nu}+\frac{1}{2}g_{\mu\nu}\lambda=0.
\end{equation}
Tracing it with $g^{\mu\nu}$ then gives
\begin{equation}
\lambda=-\frac{1}{2}G,
\end{equation}
which when substituted back yields the tracelees Einstein equation
\begin{equation}
G_{\mu\nu}-\frac{1}{4}g_{\mu\nu}G=0.
\end{equation}
Applying the Bianchi identity to this equation implies
\begin{equation}
\partial_{\mu}G=0=\partial_{\mu}\lambda,
\end{equation}
and thus
\begin{align}
\lambda &  =-4\Lambda\\
G  &  =4\Lambda,
\end{align}
where $\Lambda$ is the cosmological constant arising as an integrating
constant \cite{HT}. Therefore we see that an added benefit of having the
quantization condition is that the cosmological constant now appears as an
integrating constant in the equations of motion and is not necessary to be
present in the action. This result is similar to the one encountered in
unimodular gravity, with a major difference that in our case the
diffeomorphism symmetry is not restricted but only the volume is quantized
with all symmetries being intact.

Next, varying the fields $Y^{A}$ gives (using $\partial_{\mu}\lambda=0$ )
\begin{equation}
-\frac{5}{2(4!)}\lambda\epsilon^{\mu\nu\kappa\lambda}\epsilon_{ABCDE}%
\partial_{\mu}Y^{B}\partial_{\nu}Y^{C}\partial_{\kappa}Y^{D}\partial_{\lambda
}Y^{E}+\lambda^{\prime}Y_{A}\sqrt{g}=0.
\end{equation}
Tracing this equation with $Y^{A}$ gives%
\begin{equation}
\lambda^{\prime}=\frac{5}{2}\lambda=-\frac{5}{4}G.
\end{equation}
Assuming that $G\neq0$ (the case $G=0$ recovers the full set of Einstein
equations without cosmological constant), we further have
\begin{equation}
Y_{A}=\frac{1}{4!}\frac{1}{\sqrt{g}}\epsilon^{\mu\nu\kappa\lambda}%
\epsilon_{ABCDE}\partial_{\mu}Y^{B}\partial_{\nu}Y^{C}\partial_{\kappa}%
Y^{D}\partial_{\lambda}Y^{E}, \label{trivial}%
\end{equation}
which implies the equation
\begin{equation}
\epsilon^{\mu\nu\kappa\lambda}\epsilon_{A^{\prime}BCDE}\partial_{\mu}%
Y^{B}\partial_{\nu}Y^{C}\partial_{\kappa}Y^{D}\partial_{\lambda}Y^{E}\left(
\delta_{A}^{A^{\prime}}-Y_{A}Y^{A^{\prime}}\right)  =0.
\end{equation}
Note that the expression
\begin{align}
\frac{3}{8\pi^{2}}\frac{1}{4!}\int_{S^{4}}d^{4}x\epsilon^{\mu\nu\kappa\lambda
}\epsilon_{ABCDE}Y^{A}\partial_{\mu}Y^{B}\partial_{\nu}Y^{C}\partial_{\kappa
}Y^{D}\partial_{\lambda}Y^{E}  &  =\pi_{4}\left(  S^{4}\right) \\
&  =\mathbb{Z},
\end{align}
is the winding of the sphere $S^{4}$ ($\pi_{n}\left(  S^{n}\right)
=\mathbb{Z}$ \cite{Greub}, \cite{Bott}). Thus
\begin{equation}%
%TCIMACRO{\dint \limits_{M}}%
%BeginExpansion
{\displaystyle\int\limits_{M}}
%EndExpansion
\sqrt{g}d^{4}x=N\left(  \frac{8\pi^{2}}{3}\right)  ,
\end{equation}
where $N$ is the winding number of the mapping $M_{4}\rightarrow S^{4}$
\cite{Ram} \cite{Coleman}. We can easily see that the $Y^{A}$ equation of
motion (\ref{trivial}) follows from equation (\ref{volume}) and does not give
any new information because it appears through a topological term. To see this
use the identity resulting from anti-symmetrizing six indices taking five
values,
\begin{align}
0  &  =Y_{\left[  A\right.  }\epsilon_{\left.  A^{\prime}BCDE\right]
}\epsilon^{\mu\nu\kappa\lambda}\partial_{\mu}Y^{B}\partial_{\nu}Y^{C}%
\partial_{\kappa}Y^{D}\partial_{\lambda}Y^{E}\\
&  =\left(  Y_{A}\epsilon_{A^{\prime}BCDE}-Y_{A^{\prime}}\epsilon
_{ABCDE}-4Y_{B}\epsilon_{AA^{\prime}CDE}\right)  \epsilon^{\mu\nu\kappa
\lambda}\partial_{\mu}Y^{B}\partial_{\nu}Y^{C}\partial_{\kappa}Y^{D}%
\partial_{\lambda}Y^{E},\nonumber
\end{align}
which, after using the property $Y_{B}\partial_{\mu}Y^{B}=0$ and equation
(\ref{volume}), implies equation (\ref{trivial}).

\section{Solitonic solution}

We have seen that if we consider the spectral action to be of the form
$\mathrm{Tr}f\left(  D,Y\right)  ,$\ it will then contain the kinetic term
\begin{equation}
\frac{1}{2}%
%TCIMACRO{\dint \limits_{M}}%
%BeginExpansion
{\displaystyle\int\limits_{M}}
%EndExpansion
d^{4}x\sqrt{g}g^{\mu\nu}\partial_{\mu}Y^{A}\partial_{\nu}Y^{A}.
\end{equation}
Including this term in the action gives the modified Einstein equations
\begin{equation}
G_{\mu\nu}+\frac{1}{2}g_{\mu\nu}\lambda=\partial_{\mu}Y^{A}\partial_{\nu}%
Y^{A}-\frac{1}{2}g_{\mu\nu}\left(  \partial Y\cdot\partial Y\right)  ,
\label{Einstein}%
\end{equation}
where we have denoted by $\partial Y\cdot\partial Y=g^{\kappa\lambda}%
\partial_{\kappa}Y^{A}\partial_{\lambda}Y^{A}.$ Taking the trace of this
equation determines $\lambda:$%
\begin{equation}
\lambda=-\frac{1}{2}\left(  G+\partial Y\cdot\partial Y\right)  ,
\label{Gequation}%
\end{equation}
and when this is plugged back into equation (\ref{Einstein}) it gives two
equations, the first of which is traceless
\begin{equation}
G_{\mu\nu}-\frac{1}{4}g_{\mu\nu}G=\partial_{\mu}Y^{A}\partial_{\nu}Y^{A}%
-\frac{1}{4}g_{\mu\nu}\left(  \partial Y\cdot\partial Y\right)  .
\end{equation}
Taking covariant derivative of equation (\ref{Einstein}) using Bianchi
identity, gives%
\begin{equation}
\frac{1}{2}\partial_{\mu}\lambda=\partial_{\mu}Y^{A}\square Y^{A},
\label{lambda}%
\end{equation}
where $\square Y^{A}=g^{\mu\nu}\nabla_{\mu}\partial_{\nu}Y^{A}$ and after
making use of the identity $Y^{A}\square Y^{A}=-\partial Y\cdot\partial Y$
that follows by differentiating $Y^{A}\partial_{\mu}Y^{A}=0.$ We now examine
the $Y^{A}$ equation
\begin{align}
&  -\frac{5}{2(4!)}\lambda\epsilon^{\mu\nu\kappa\lambda}\epsilon
_{ABCDE}\partial_{\mu}Y^{B}\partial_{\nu}Y^{C}\partial_{\kappa}Y^{D}%
\partial_{\lambda}Y^{E}+\lambda^{\prime}Y_{A}\sqrt{g}\nonumber\\
&  =\sqrt{g}\square Y^{A}+\frac{1}{12}\epsilon^{\mu\nu\kappa\lambda}%
\epsilon_{ABCDE}\partial_{\mu}\lambda Y^{B}\partial_{\nu}Y^{C}\partial
_{\kappa}Y^{D}\partial_{\lambda}Y^{E}.
\end{align}
Tracing with $Y^{A}$ gives%
\begin{equation}
\lambda^{\prime}=\frac{5}{2}\lambda+Y^{A}\square Y^{A}.
\end{equation}
Plugging this back and using equation (\ref{trivial} ) gives
\begin{equation}
\square Y^{A}-Y^{A}\left(  Y^{B}\square Y^{B}\right)  =-\frac{1}{12\sqrt{g}%
}\epsilon^{\mu\nu\kappa\lambda}\epsilon_{ABCDE}\partial_{\mu}\lambda
Y^{B}\partial_{\nu}Y^{C}\partial_{\kappa}Y^{D}\partial_{\lambda}Y^{E}.
\label{Yequation}%
\end{equation}
The left-hand side of equation (\ref{lambda}) is a total derivative, while the
right-hand side is not. The general solution of equations (\ref{lambda}) and
(\ref{Yequation}) is not easy to find. We shall restrict ourselves to the
subspace where
\[
\partial_{\mu}\lambda=0,
\]
so that
\[
G+g^{\mu\nu}\partial_{\mu}Y^{A}\partial_{\nu}Y^{A}=4\Lambda.
\]
Equation (\ref{Yequation}) then simplifies to
\begin{equation}
\square Y^{A}-Y^{A}\left(  Y^{B}\square Y^{B}\right)  =0. \label{yequation}%
\end{equation}
This equation, being traceless, could be recast in terms of the dependent
variables $Y^{a},$ $a=1,\cdots4,$ substituting the relation $Y^{5}%
=\sqrt{1-Y^{a}Y^{a}}$ so that the kinetic term $g^{\mu\nu}\partial_{\mu}%
Y^{A}\partial_{\nu}Y^{A}$ takes the form
\begin{equation}
g^{\mu\nu}\partial_{\mu}Y^{a}\partial_{\nu}Y^{b}h_{ab},,
\end{equation}
where
\begin{equation}
h_{ab}=\left(  \delta_{ab}+\frac{Y_{a}Y_{b}}{1-Y^{c}Y^{c}}\right)  .
\end{equation}
The equation (\ref{yequation}) then takes the form \cite{Xin}
\begin{equation}
g^{\mu\nu}\left(  \nabla_{\mu}\partial_{\nu}Y^{a}+\partial_{\mu}Y^{b}%
\partial_{\nu}Y^{c}\Gamma_{bc}^{a}\right)  =0,
\end{equation}
where $\Gamma_{bc}^{a}$ is the Christoffel connection of the metric $h_{ab}$
on the sphere $S^{4}$ which is given by
\begin{equation}
\Gamma_{bc}^{a}=h_{bc}Y^{a}.
\end{equation}
This shows that the fields $Y^{a}$ are harmonic maps which shows that maps
from the four-manifolds $M_{4}$ to $S^{4}$ satisfying the equations of motion
are harmonic. We conclude that the equations of motion are identical to those
of the $O(5)$ non-linear sigma model, which is also equivalent to the
Projective quaternionic model $HP^{1}$ \cite{Gava}, \cite{Gursey}. These works
have derived the instanton solution (for a conformally flat metric) with $N=1$
and the multi-instanton solution $N=n$.

First for the $N=1$ instanton solution we have%
\begin{align}
g_{\mu\nu}  &  =\delta_{\mu\nu}\frac{1}{\left(  1+x^{2}\right)  ^{2}},\quad
x^{2}=x^{a}x^{a},\qquad a=1,\cdots,4\\
Y^{a}  &  =\frac{2x^{a}}{1+x^{2}},\quad Y^{5}=\frac{x^{2}-1}{1+x^{2}},
\end{align}
which satisfies
\begin{equation}
R_{\mu\nu}=\frac{1}{4}g_{\mu\nu}R,\quad R=48.
\end{equation}
The multi-instanton solution is given by
\begin{equation}
g_{\mu\nu}=2\left(  \partial_{\mu}x^{n}\partial_{\nu}\overline{x}^{n}%
+\partial_{\nu}x^{n}\partial_{\mu}\overline{x}^{n}\right)  \frac{1}{\left(
1+x^{n}\overline{x}^{n}\right)  ^{2}},
\end{equation}
where $x$ is a quaternionic coordinate%
\begin{equation}
x=x^{4}1+e_{i}x^{i},
\end{equation}
where $e_{i}$, $i=1,2,3$ are the three quaternionic complex structures
$e_{1}^{2}=e_{2}^{2}=e_{3}^{2}=-1$ and $e_{1}e_{2}=-e_{2}e_{1}=e_{3}.$ We also
have
\begin{align}
Y  &  =\frac{2x^{n}}{\left(  1+x^{n}\overline{x}^{n}\right)  }=Y^{4}%
1+e^{i}Y^{i}\\
Y^{5}  &  =\frac{\left(  x^{n}\overline{x}^{n}-1\right)  }{\left(
x^{n}\overline{x}^{n}+1\right)  }.
\end{align}
This solution gives a winding number $n.$

\section{Three dimensional volume quantization}

\bigskip Up to this point we have been dealing with compact manifolds.
Physical space-time has a Lorentzian signature, and is thus topologically
equivalent to $\mathbb{R}\times M_{3}.$

Alternatively, we can envision the following picture. Consider as a starting
point any three dimensional hypersurface $\Sigma_{3}$ whose normals at any
point has time-like directions and with a family of geodesic lines normal to
the hypersurface. Let these lines be time coordinates and set $t$ to be the
distance as measured from the initial hypersurface. Denote by $y^{i},$
$i=1,2,3$ as the coordinates on the hypersurface $\Sigma.$ There will still be
arbitrary coordinate transformations $x^{\alpha}=x^{\alpha}\left(
y^{\emph{i}}\right)  ,$ $\alpha=1,2,3.$ Denote the four coordinates by
$x^{\mu}=\left(  t,x^{\alpha}\right)  $ and define the functions \cite{Kuchar}%
\begin{equation}
e_{i}^{\mu}=\frac{\partial x^{\mu}}{\partial y^{i}},
\end{equation}
and the corresponding normal vectors $n_{\mu}$ such that
\begin{equation}
n_{\mu}e_{i}^{\mu}=0.
\end{equation}
The inverse functions $e_{\mu}^{i}$ are defined with the aid of the vectors
$n_{\mu}$ so that%
\begin{equation}
e_{i}^{\mu}e_{\mu}^{j}=\delta_{i}^{j},\qquad e_{i}^{\mu}e_{\nu}^{i}%
=\delta_{\nu}^{\mu}-n^{\mu}n_{\nu},
\end{equation}
where the vectors $n_{\mu}$ satisfy
\begin{equation}
n_{\mu}n^{\mu}=\varepsilon,
\end{equation}
where $\varepsilon=1$ for metric with signature $\left(  +,+,+,+\right)  $ and
$\varepsilon=-1$ for signature $\left(  -,+,+,+\right)  .$ The metric on the
four-dimensional manifold generated due to the motion of the three dimensional
hypersurface is then given by%
\begin{equation}
g_{\mu\nu}=e_{\mu}^{i}h_{ij}e_{\nu}^{j}+\varepsilon n_{\mu}n_{\nu},
\end{equation}
where $h_{ij}$ is the metric on the three dimensional hypersurface $\Sigma.$
The inverse metric is given by%
\begin{equation}
g^{\mu\nu}=e_{i}^{\mu}h^{ij}e_{j}^{\nu}+\varepsilon n^{\mu}n^{\nu},
\end{equation}
where $h^{ij}$ is the inverse metric of $h_{ij}$ which implies that
\begin{equation}
n^{\mu}=g^{\mu\nu}n_{\nu},\qquad n^{\mu}e_{\mu}^{i}=0.
\end{equation}
For simplicity we can chose the gauge where
\begin{equation}
e_{i}^{t}=\frac{\partial t}{\partial y^{i}}=0,
\end{equation}
which implies that
\begin{equation}
n_{\alpha}=0.
\end{equation}
Denoting
\begin{equation}
n_{t}=N,\qquad e_{t}^{i}=N^{i},
\end{equation}
the components of the metric $g_{\mu\nu}$ will be given by
\begin{equation}
g_{tt}=\varepsilon N^{2}+N^{\alpha}h_{\alpha\beta}N^{\beta},\qquad g_{t\alpha
}=N_{\alpha},\qquad g_{\alpha\beta}=h_{\alpha\beta},
\end{equation}
where
\begin{equation}
h_{\alpha\beta}=e_{\alpha}^{i}h_{ij}e_{\beta}^{j},\qquad N_{\alpha}=e_{\alpha
}^{i}h_{ij}N^{j}.
\end{equation}
In particular, the vector $n^{\mu}$ is given by
\begin{equation}
n^{\mu}=\left(  \frac{1}{N},-\frac{N^{\alpha}}{N}\right)  .
\end{equation}
This gives the familiar $3+1$ ADM splitting of the metric \cite{MTW}
\begin{equation}
ds^{2}=h_{\alpha\beta}\left(  dx^{\alpha}+N^{\alpha}dt\right)  \left(
dx^{\beta}+N^{\beta}dt\right)  +\varepsilon N^{2}dt^{2}.
\end{equation}
At this point we note that for the three dimensional hypersurface $\Sigma_{3}$
we will utilize the two maps $Y$ and $Y^{\prime}$ from $\Sigma$ to the three
sphere $S^{3},$ which are defined with respect to the Clifford algebras
$\mathrm{Cliff}\left(  +,+,+,+\right)  =M_{2}\left(  \mathbb{H}\right)  $ and
$\mathrm{Cliff}\left(  -,-,-,-\right)  =M_{2}\left(  \mathbb{H}\right)  $
where%
\begin{equation}
Y=Y^{a}\Gamma_{a},\qquad Y^{\prime}=iY^{\prime a}\Gamma_{a}^{\prime},\qquad
a=1,\cdots,4,
\end{equation}
where
\begin{equation}
\left\{  \Gamma_{a},\Gamma_{b}\right\}  =2\delta_{ab},\qquad\left\{
\Gamma_{a}^{\prime},\Gamma_{b}^{\prime}\right\}  =-2\delta_{ab},
\end{equation}
and $Y^{2}=1,$ $Y^{\prime2}=1.$ In reality, we can consider the mappings from
the moving hypersurfaces $\Sigma_{3}$ which generate the four dimensional
manifold and thus we have $Y^{a}\left(  x^{\mu}\right)  $ and $Y^{\prime
a}\left(  x^{\mu}\right)  .$ These could be extended by the field $X\left(
x^{\mu}\right)  $ which maps the geodesics normal to $\Sigma_{3}$ into
$\mathbb{R}.$ We can then consider the field $X$ to be measure of the
distance
\begin{equation}
X=\sqrt{g_{\mu\nu}dx^{\mu}dx^{\nu}},
\end{equation}
which according to the Hamilton-Jacobi equation will then satisfy \cite{LL}%
\begin{equation}
g^{\mu\nu}\frac{\partial X}{\partial x^{\mu}}\frac{\partial X}{\partial
x^{\nu}}=\varepsilon, \label{mimetic}%
\end{equation}
and this is a requirement that the mapping function $X$ preserves the length
of a curve on $M_{4}.$ This relation could be viewed as a condition to
minimize the distance between two points in noncommutative geometry%
\begin{equation}
\left[  D,X\right]  ^{2}=-1.
\end{equation}
Thus, in contrast to the four dimensional case where the mapping is from
$M_{4}$ to $S^{4}\times S^{4},$ the mapping now is from $\mathbb{R}%
\times\Sigma_{3}$ to $\mathbb{R}\times S^{3}\times S^{3}.$ The Feynman slashed
fields $Y^{5}\Gamma_{5}$ and $Y^{\prime5}\Gamma_{5}^{\prime}$ must now be
replaced with the field $X$ slashed with some combination of $1,$ $\Gamma
_{5},$ $\Gamma_{5}^{\prime}$ and $\Gamma_{5}\Gamma_{5}^{\prime}.$ To find out
the correct procedure, we make the following observation. In the
four-dimensional case, we used the Feynman slashed coordinates $Y=Y^{A}%
\Gamma_{A},$ $A=1,\cdots,5.$ The matrices $\frac{1}{4}\Gamma_{AB}=\frac{1}%
{8}\left(  \Gamma_{A}\Gamma_{B}-\Gamma_{B}\Gamma_{A}\right)  $ are generators
of the Lie Algebra $SO\left(  5\right)  .$ Denoting these by $J_{AB}$, they
have the commutation relations
\begin{equation}
\left[  J_{AB},J_{CD}\right]  =-\left(  \delta_{AC}J_{BD}-\delta_{BC}%
J_{AD}-\delta_{AD}J_{BC}+\delta_{BD}J_{AC}\right)  .
\end{equation}
Denoting $A=a,5$ where $a=1,\cdots,4$ and $J_{a5}=RP_{a}$ we then have
\begin{equation}
\left[  P_{a},P_{b}\right]  =-\frac{1}{R^{2}}J_{ab}.
\end{equation}
In the limit $R=\frac{1}{\eta}\rightarrow\infty,$ the generators $P_{a}$
become, locally, the translation generators and $J_{ab}$ will correspond to
$SO\left(  4\right)  $ Lorentz generators. This is the procedure we will
follow to decompose one of the coordinates, say $Y^{5}$ by writing
\begin{equation}
Y^{5}=\eta X,
\end{equation}
and simultaneously rescale one of the coordinates, say $x^{4}$
\begin{equation}
x^{4}\rightarrow\eta t,
\end{equation}
then taking the limit $\eta\rightarrow0.$ We will obtain the volume
quantization condition by compactifying the four-dimensional two sided
relation to $3+1$ in the above limit, where the fields $Y^{5}$ and
$Y^{\prime5}$ are not coordinates on the fours sphere, but independent fields.
To this end, let
\begin{align}
Z  &  =2EE^{\prime}-1\\
&  =\frac{1}{2}\left(  Y^{a}\Gamma_{a}+\eta X\Gamma_{5}+1\right)  \left(
Y^{^{\prime}a}\Gamma_{a}+\eta X\Gamma_{5}^{\prime}+1\right)  -1\\
&  =2ee^{\prime}-1+\eta X\left(  \Gamma_{5}e^{\prime}+\Gamma_{5}^{\prime
}e\right)  +O\left(  \eta^{2}\right) \\
&  =z+\eta X\left(  \Gamma_{5}e^{\prime}+\Gamma_{5}^{\prime}e\right)
+O\left(  \eta^{2}\right)  ,
\end{align}
where
\begin{equation}
e=\frac{1}{2}\left(  Y^{a}\Gamma_{a}+1\right)  ,\qquad e^{\prime}=\frac{1}%
{2}\left(  Y^{\prime a}\Gamma_{a}^{\prime}+1\right)  ,\qquad z=2ee^{\prime}-1.
\end{equation}
Notice that we have identified the fields $Y^{5}$ and $Y^{\prime5}$ with the
same field $X$ because this is the field corresponding to the motion of the
hypersurface. The correct quantization condition of the $3+1$ dimensional
space, which also results from compactification of the four dimensional
quantization condition is given by
\begin{equation}
\lim_{\eta\rightarrow0}\frac{1}{\eta}\left\langle \left(  z+\eta X\left(
\Gamma_{5}e^{\prime}+\Gamma_{5}^{\prime}e\right)  \right)  \left(  \left[
D,z\right]  +\eta\left[  D,X\left(  \Gamma_{5}e^{\prime}+\Gamma_{5}^{\prime
}e\right)  \right]  \right)  ^{4}\right\rangle =\gamma,
\end{equation}
where $\gamma$ is the chirality operator of the generated $3+1$ dimensional
manifold. For consistency, one must first show that all terms of order
$\frac{1}{\eta}$ are zero. For example%
\begin{equation}
\left\langle z\left[  D,z\right]  ^{4}\right\rangle =0,
\end{equation}
as this would involve terms like $\left\langle \Gamma_{a}\Gamma_{b}\Gamma
_{c}\Gamma_{d}\Gamma_{e}\right\rangle =0$ because this is the trace of an odd
number of $\Gamma$ matrices. Therefor we have to worry only about terms
independent of $\eta$ as the terms of order $\eta$ vanish in the limit. Terms
which are linear in $X$ (and not its derivative) also vanish because terms of
the form
\begin{equation}
X\left\langle \left(  \Gamma_{5}e^{\prime}+e\Gamma_{5}^{\prime}\right)
\left[  D,z\right]  ^{4}\right\rangle ,
\end{equation}
will give the terms
\begin{equation}
X\epsilon^{\mu\nu\kappa\lambda}\epsilon_{abcd}\partial_{\mu}Y^{a}\partial
_{\nu}Y^{b}\partial_{\kappa}Y^{c}\partial_{\lambda}Y^{d}=X\det\left\vert
\partial_{\mu}Y^{a}\right\vert =0,
\end{equation}
as the Jacobian $\left\vert \partial_{\mu}Y^{a}\right\vert $ vanishes because
the four $Y^{a}$ are not independent. After some algebra, one can check that
the only non-vanishing terms are
\begin{equation}%
%TCIMACRO{\dsum \limits_{p=0}^{3}}%
%BeginExpansion
{\displaystyle\sum\limits_{p=0}^{3}}
%EndExpansion
\left\langle z\left[  D,z\right]  ^{p}\left(  \left[  D,X\right]  \right)
\left(  \Gamma_{5}e^{\prime}+e\Gamma_{5}^{\prime}\right)  \left[  D,z\right]
^{3-p}\right\rangle =\gamma. \label{3dtwosided}%
\end{equation}
There is no need to repeat the calculation done in the $d=4$ case as the
result holds in general, and in particular in the limit $\eta\rightarrow0$ and
this is a smooth limit as terms of order $\frac{1}{\eta}$ vanish identically.
We thus conclude that this condition implies
\begin{equation}
\frac{1}{3!}\epsilon^{\mu\nu\kappa\lambda}\epsilon_{abcd}\partial_{\mu
}X\left(  Y^{a}\partial_{\nu}Y^{b}\partial_{\kappa}Y^{c}\partial_{\lambda
}Y^{d}+Y^{^{\prime}a}\partial_{\nu}Y^{\prime b}\partial_{\kappa}Y^{\prime
c}\partial_{\lambda}Y^{\prime d}\right)  =\det\left\vert e_{\mu}%
^{a}\right\vert .
\end{equation}
The field $X$ could be identified with the time coordinate in a certain gauge.
For example, in the synchronous gauge we have $g^{tt}=1,$ $g^{ti}=0$ which
implies that $X=t$ is a solution of the above constraint. If we define the
three-dimensional hypersurface $\Sigma_{3}$ by $t=$constant, then the lapse
function $N$ could be defined by $\partial_{t}X=N$ with the boundary
condition
\begin{equation}
\partial_{i}X|_{\Sigma}=0.
\end{equation}
We could have obtained the $3+1$ quantization condition, directly by
compactifying the four-dimensional condition of the mapping from
$M_{4}\rightarrow S^{4}.$ Let $Y^{5}=\eta X=Y^{\prime5}$ and simultaneously
rescale one of the coordinates, say $x^{4}$
\begin{equation}
x^{4}\rightarrow\eta x^{0},
\end{equation}
so that the constraint in the limit $\eta\rightarrow0,$ becomes (written
covariantly)
\begin{align}
\sqrt{g}  &  =\lim_{\eta\rightarrow0}\left(  \frac{1}{4!}\epsilon^{\mu
\nu\kappa\lambda}\epsilon_{ABCDE}\left(  Y^{A}\partial_{\mu}Y^{B}\partial
_{\nu}Y^{C}\partial_{\kappa}Y^{D}\partial_{\lambda}Y^{E}\right.  \right. \\
&  \qquad\qquad\left.  \left.  +Y^{\prime A}\partial_{\mu}Y^{\prime B}%
\partial_{\nu}Y^{^{\prime}C}\partial_{\kappa}Y^{^{\prime}D}\partial_{\lambda
}Y^{^{\prime}E}\right)  \right) \\
&  =\frac{1}{3!}\epsilon^{\mu\nu\kappa\lambda}\epsilon_{abcd}\left(
\partial_{\mu}X\right)  \left(  Y^{a}\partial_{\nu}Y^{b}\partial_{\kappa}%
Y^{c}\partial_{\lambda}Y^{d}+Y^{^{\prime}a}\partial_{\nu}Y^{^{\prime}%
b}\partial_{\kappa}Y^{^{\prime}c}\partial_{\lambda}Y^{\prime d}\right)  ,
\end{align}
where the $X$ field is unconstrained, while the fields $Y^{a}$ and
$Y^{^{\prime}a}$ satisfy
\begin{equation}
Y^{a}Y^{a}=1,\qquad Y^{^{\prime}a}Y^{^{\prime}a}=1,\qquad a=1,\cdots,4.
\end{equation}
Notice that the term
\begin{equation}
\frac{1}{4!}\epsilon^{\mu\nu\kappa\lambda}X\partial_{\mu}Y^{a}\partial_{\nu
}Y^{b}\partial_{\kappa}Y^{c}\partial_{\lambda}Y^{d}\epsilon_{abcd}%
=XdY^{1}\wedge dY^{2}\wedge dY^{3}\wedge dY^{4},
\end{equation}
is equal to zero because $dY^{4}$ depends on a linear combination of
$dY^{1},\cdots,dY^{3}.$

The Clifford algebra $M_{2}\left(  \mathbb{H}\right)  \oplus M_{2}\left(
\mathbb{H}\right)  $ spanned by $Y^{a}\Gamma_{a}$ and $Y^{^{\prime}a}%
\Gamma_{a}^{\prime}$ will be extended by the generators $X\Gamma_{5}$ and
$X\Gamma_{5}^{\prime}.$ The first $M_{2}\left(  \mathbb{H}\right)  $
corresponding to the Clifford algebra $\mathrm{Cliff}(+,+,+,+)$ is not
effected by the addition of $\Gamma_{5}.$ The second $M_{2}\left(
\mathbb{H}\right)  $ corresponding to the Clifford algebra $\mathrm{Cliff}%
(-,-,-,-)$ changes to $M_{4}\left(  \mathbb{C}\right)  $ when extended by
$\Gamma_{5}^{\prime}.$ Thus the algebra associated with the two sided relation
(\ref{3dtwosided}) for the $3+1$ manifold is the same as the four dimensional
case and is given by
\begin{equation}
M_{2}\left(  \mathbb{H}\right)  \oplus M_{4}\left(  \mathbb{C}\right)  .
\end{equation}
Thus, this compactification corresponds to a mapping from $\mathbb{R}%
\times\Sigma_{3}\rightarrow\mathbb{R\times}S^{3}$ where $\Sigma_{3}$ is a
three dimensional hypersurface. Although imposing this condition could be made
and leads to the mimetic matter phenomena \cite{CM},\cite{CMV}, it is worth
noting that we need to impose this condition only on the hypersurface
$\Sigma_{3}$ to be defined below:
\begin{equation}
g^{\mu\nu}\partial_{\mu}X\partial_{\nu}X|_{\Sigma}=1.
\end{equation}
To get acquainted with this condition, we first consider the situation where
we have a three dimensional hypersurface in space-time, a case dealt with in
the ADM decomposition \cite{MTW}. Consider the $3+1$ splitting of space-time
so that (for Lorentzian signature)
\begin{equation}
ds^{2}=h_{ij}\left(  dx^{i}+N^{i}dt\right)  \left(  dx^{j}+N^{j}dt\right)
-N^{2}dt^{2},
\end{equation}
where $N\left(  x^{i},t\right)  $ and $N^{i}\left(  x^{i},t\right)  $ are the
lapse and shift functions. Then
\begin{equation}
\sqrt{-g}=N\sqrt{h}.
\end{equation}
We, therefore, supplement the volume quantization condition
\begin{equation}
\sqrt{g}=\frac{1}{3!}\epsilon^{\mu\nu\kappa\lambda}\epsilon_{abcd}%
\partial_{\mu}X\left(  Y^{a}\partial_{\nu}Y^{b}\partial_{\kappa}Y^{c}%
\partial_{\lambda}Y^{d}+Y^{^{\prime}a}\partial_{\nu}Y^{^{\prime}b}%
\partial_{\kappa}Y^{^{\prime}c}\partial_{\lambda}Y^{^{\prime}d}\right)  ,
\label{volume3}%
\end{equation}
by adding the constraints (\ref{mimetic}) to hold on the hypersurface
\begin{equation}
\partial_{i}X|_{\Sigma}=0,\qquad\partial_{t}X|_{\Sigma}=N|_{\Sigma},
\end{equation}
from which we deduce that the constraint (\ref{volume3}), when restricted to
the hypersuface $\Sigma_{3},$ gives
\begin{equation}
\left(  N\sqrt{h}\right)  _{\Sigma}=\frac{1}{3!}N\epsilon^{ijk}\epsilon
_{abcd}\left(  Y^{a}\partial_{i}Y^{b}\partial_{j}Y^{c}\partial_{k}%
Y^{d}+Y^{^{\prime}a}\partial_{i}Y^{^{\prime}b}\partial_{j}Y^{^{\prime}%
c}\partial_{k}Y^{^{\prime}d}\right)  ,
\end{equation}
and we finally have
\begin{align}%
%TCIMACRO{\dint \limits_{\Sigma}}%
%BeginExpansion
{\displaystyle\int\limits_{\Sigma}}
%EndExpansion
\sqrt{h}d^{3}x  &  =\frac{1}{3!}%
%TCIMACRO{\dint \limits_{\Sigma}}%
%BeginExpansion
{\displaystyle\int\limits_{\Sigma}}
%EndExpansion
\epsilon^{ijk}\epsilon_{abcd}\left(  Y^{a}\partial_{i}Y^{b}\partial_{j}%
Y^{c}\partial_{k}Y^{d}+Y^{^{\prime}a}\partial_{i}Y^{^{\prime}b}\partial
_{j}Y^{^{\prime}c}\partial_{k}Y^{^{\prime}d}\right)  d^{3}x\\
&  =\frac{1}{3!}%
%TCIMACRO{\dint \limits_{\Sigma}}%
%BeginExpansion
{\displaystyle\int\limits_{\Sigma}}
%EndExpansion
\epsilon_{abcd}\left(  Y^{a}dY^{b}dY^{c}dY^{d}+Y^{^{\prime}a}dY^{\prime
b}dY^{^{\prime}c}dY^{^{\prime}d}\right) \\
&  =\frac{4}{3}\pi^{2}\left(  w+w^{\prime}\right)
\end{align}
where $w$ and $w^{\prime}$ are integers given by the winding numbers on
$S^{3}.$ One can check that an exact solitonic solution with winding number
one, is given by
\begin{equation}
X=t,\qquad Y^{m}=\frac{2x^{m}}{1+x^{m}x^{m}},\qquad Y^{4}=\frac{x^{m}x^{m}%
-1}{1+x^{m}x^{m}},
\end{equation}
with the metric
\begin{equation}
g_{tt}=1,\qquad g_{t\alpha}=0,\qquad g_{\alpha\beta}=\frac{\delta_{\alpha
\beta}}{\left(  1+x^{m}x^{m}\right)  ^{2}},
\end{equation}
and this corresponds to a quantized three dimensional volume.

To understand the condition $g^{\mu\nu}\partial_{\mu}X\partial_{\nu}%
X|_{\Sigma}=1$ we notice that in the synchronous gauge \cite{LL} we can take
$X=\tau,$ $g^{tt}=\frac{1}{N^{2}}$ so that $\left\vert \frac{\partial\tau
}{\partial t}\right\vert =N$ and thus the line measure $Ndt\rightarrow
N\frac{\partial t}{\partial\tau}d\tau=d\tau$ which is consistent with
$g^{\tau\tau}=1.$ Thus this condition amounts to length preserving
transformation. We deduce that in a Lorentzian space-time volume quantization
is possible, provided that the field corresponding to the non-compact
transformation satisfy a length preserving condition. For the two sided
equation where we have both $Y^{A}$ and $Y^{\prime A}$ it is important to
truncate both $Y^{5}$ and $Y^{\prime5}$ to the same field $X$%
\[
Y^{5}=\eta X,\qquad Y^{\prime5}=\eta X,
\]
which avoids imposing further unnatural conditions. There are many advantages
to impose the condition (\ref{mimetic}) locally as this constraint modifies
Einstein gravity only in the longitudinal sector as the field $X$ is not
dynamical. In the synchronous gauge, this field is identified with the time
coordinate and modifies Einstein equations by giving an energy-momentum tensor
in the absence of matter, giving rise to mimetic cold matter. We have shown
that this field, which arises naturally from the three space quantization
condition can be used to construct realistic cosmological models such as
inflation without the need to introduce additional scalar fields. By including
terms in the action of the form $f\left(  \square X\right)  $ which do occur
in the spectral action as can be seen from considerations of the scale
invariance, it is possible to avoid singularities in Friedmann, Kasner
\cite{CMK} or Black hole solutions \cite{CMBH}. This is possible because the
contributions of the field $X$ to the energy-momentum tensor would allow, for
special functions $f\left(  \square X\right)  $ to limit the curvature,
preventing the singularities from occurring.

\section{Area quantization}

\bigskip Next consider the compactification of two fields, keeping only three
compact fields $Y^{m},$ $m=1,2,3,$ and rescale the two fields
\begin{align}
Y^{4}  &  =\eta X^{1},\qquad Y^{5}=\eta X^{2}\\
Y^{\prime4}  &  =\eta X^{1},\qquad Y^{^{\prime}5}=\eta X^{2},
\end{align}
and simultaneously rescale the coordinates
\begin{equation}
x^{\alpha}\rightarrow\eta x^{\alpha},\text{\qquad}\alpha=1,2,
\end{equation}
where $x^{\alpha}$ are coordinates along directions transverse to the two
dimensional hypersurface, so that
\begin{align}
\sqrt{g}  &  =\lim_{\eta\rightarrow0}\left(  \frac{1}{4!}\epsilon^{\mu
\nu\kappa\lambda}\epsilon_{ABCDE}\left(  Y^{A}\partial_{\mu}Y^{B}\partial
_{\nu}Y^{C}\partial_{\kappa}Y^{D}\partial_{\lambda}Y^{E}\right.  \right. \\
&  \qquad\qquad\left.  \left.  +Y^{^{\prime}A}\partial_{\mu}Y^{^{\prime}%
B}\partial_{\nu}Y^{^{\prime}C}\partial_{\kappa}Y^{^{\prime}D}\partial
_{\lambda}Y^{^{\prime}E}\right)  \right) \\
&  =\frac{1}{2}\epsilon^{\mu\nu\kappa\lambda}\epsilon_{ab}\partial_{\mu}%
X^{a}\partial_{\nu}X^{b}\epsilon_{mnp}\left(  Y^{m}\partial_{\kappa}%
Y^{n}\partial_{\lambda}Y^{p}+Y^{^{\prime}m}\partial_{\kappa}Y^{^{\prime}%
n}\partial_{\lambda}Y^{\prime p}\right)  , \label{volume2}%
\end{align}
where $X^{a}\left(  x^{\mu}\right)  ,$ $a=1,2,$ while the $Y^{m}\left(
x^{\mu}\right)  $ and $Y^{\prime m}\left(  x^{\mu}\right)  $ are subject to
the constraints%
\begin{equation}
Y^{p}Y^{p}=1,\qquad Y^{\prime p}Y^{^{\prime}p}=1,\qquad p=1,2,3.
\end{equation}
Again, since the functions $X^{a}$ are unconstrained to be coordinates on a
sphere, normalization conditions must be imposed
\begin{equation}
\det\left(  g^{\mu\nu}\partial_{\mu}X^{a}\partial_{\nu}X^{b}\right)  _{\Sigma
}=1. \label{normal3}%
\end{equation}
In case of Minkowski signature we must replace $1$ with $-1.$ It is known that
this condition is the area preserving transformation on the two dimensional
surface from the original surface with coordinates $x^{\alpha}$ to the surface
with coordinates $X^{a}.$ We note that in order to completely characterize
this transformation we still have the option of specifying the\textit{ trace}
of the matrix $g^{\mu\nu}\partial_{\mu}X^{a}\partial_{\nu}X^{b}$ which turns
out to determine the stability of the map under linear perturbations
\cite{Mac}.

Thus this compactification corresponds to the mapping $M_{4}\rightarrow
\mathbb{R}^{2}\mathbb{\times}S^{2}.$ We assume that there is a hypersurface
$\Sigma_{2}$ endowed with an induced metric and with coordinates $x^{i}$ so
that the four dimensional metric can be written in the form%
\begin{equation}
ds^{2}=h_{ij}\left(  dx^{i}+h^{ik}N_{i\alpha}dx^{\alpha}\right)  \left(
dx^{j}+h^{jl}N_{l\beta}dx^{\beta}\right)  +k_{\alpha\beta}dx^{\alpha}%
dx^{\beta},
\end{equation}
where $h^{ij}$ is the inverse of $h_{ij}$, the metric on $\Sigma_{2}$ with
$i,j=1,2$ and $\alpha,\beta=3,4.$ In matrix form, the four-metric is
\begin{equation}
\left(
\begin{array}
[c]{cc}%
k_{\alpha\beta}+N_{i\alpha}N_{j\beta}h^{ij} & N_{i\alpha}\\
N_{i\alpha} & h_{ij}%
\end{array}
\right)  .
\end{equation}
The inverse of this metric is given by
\begin{equation}
\left(
\begin{array}
[c]{cc}%
k^{\alpha\beta} & -N^{j\alpha}\\
-N^{j\alpha} & h^{ij}+N^{i\alpha}N^{j\beta}k_{\alpha\beta}%
\end{array}
\right)  ,
\end{equation}
where $k^{\alpha\beta}$ is the inverse of $k_{\alpha\beta}$ and $N^{i\alpha}$
is obtained from $N_{i\alpha}$ by raising indices with the metrics $h^{ij}$
and $k^{\alpha\beta}.$ The hypersurface $\Sigma_{2}$ is then defined by the
equations
\begin{equation}
x^{\alpha}=\mathrm{const,\qquad}\alpha=1,2,
\end{equation}
parametrized by the coordinates $x^{i},$ $i=3,4.$ In this form we have
\begin{equation}
\sqrt{g}=\sqrt{h}\sqrt{k}.
\end{equation}
The constraint (\ref{normal3}) is then solved by
\begin{equation}
\qquad\partial_{i}X^{a}|_{\Sigma}=0,
\end{equation}
so that
\begin{equation}
\det\left(  k^{\alpha\beta}\partial_{\alpha}X^{a}\partial_{\beta}X^{b}\right)
_{\Sigma}=1,
\end{equation}
which implies
\begin{equation}
\left(  \det k\right)  _{\Sigma}=\left(  \det\left\vert \partial_{\alpha}%
X^{a}\right\vert _{\Sigma}\right)  ^{2}.
\end{equation}
Using
\begin{align}
\left(  \epsilon^{ij}\epsilon_{ab}\epsilon^{\alpha\beta}\partial_{\alpha}%
X^{a}\partial_{\beta}X^{b}\epsilon_{mnp}Y^{m}\partial_{i}Y^{n}\partial
_{j}Y^{p}\right)  _{\Sigma}  &  =\det\left\vert \partial_{\alpha}%
X^{a}\right\vert _{\Sigma}\left(  \epsilon^{ij}\epsilon_{mnp}Y^{m}\partial
_{i}Y^{n}\partial_{j}Y^{p}\right)  _{\Sigma}\\
&  =\left(  \sqrt{k}\epsilon^{ij}\epsilon_{mnp}Y^{m}\partial_{i}Y^{n}%
\partial_{j}Y^{p}\right)  _{\Sigma}.
\end{align}
The volume constraint becomes%
\begin{equation}
\left(  \sqrt{h}\sqrt{k}\right)  _{\Sigma}=\frac{1}{2}\left(  \sqrt{k}%
\epsilon^{ij}\epsilon_{mnp}\left(  Y^{m}\partial_{i}Y^{n}\partial_{j}%
Y^{p}+Y^{^{\prime}m}\partial_{i}Y^{^{\prime}n}\partial_{j}Y^{^{\prime}%
p}\right)  \right)  _{\Sigma}.
\end{equation}
One important point to realize is that the fundamental constraint equation is
(\ref{volume2}), and that we can integrate this equation over any hypersurface
we like, and not only over the full space. In particular, let us choose to
integrate over a two dimensional hypersurface $\Sigma_{2}$ with coordinates
$x^{\alpha}$, then this implies that
\begin{align}%
%TCIMACRO{\dint \limits_{\Sigma_{2}}}%
%BeginExpansion
{\displaystyle\int\limits_{\Sigma_{2}}}
%EndExpansion
d^{2}x\sqrt{h}  &  =\frac{1}{2}%
%TCIMACRO{\dint \limits_{\Sigma}}%
%BeginExpansion
{\displaystyle\int\limits_{\Sigma}}
%EndExpansion
\epsilon^{ij}\epsilon_{mnp}\left(  Y^{m}\partial_{i}Y^{n}\partial_{j}%
Y^{p}+Y^{^{\prime}m}\partial_{i}Y^{\prime n}\partial_{j}Y^{^{\prime}p}\right)
dx^{i}dx^{j}\\
&  =\frac{1}{2}%
%TCIMACRO{\dint \limits_{\Sigma}}%
%BeginExpansion
{\displaystyle\int\limits_{\Sigma}}
%EndExpansion
\epsilon_{mnp}\left(  Y^{m}dY^{n}dY^{p}+Y^{^{\prime}m}dY^{\prime
n}dY^{^{\prime}p}\right) \\
&  =4\pi\left(  w+w^{\prime}\right)  ,
\end{align}
where $w$ and $w^{\prime}$are integers and equal to the winding numbers of the
two maps.

\section{Equations of motion for $\mathbb{R}\times S^{3}$ and $\mathbb{R}%
^{2}\times S^{2}$}

\subsection{\bigskip$\mathbb{R}\times S^{3}$ case}

Start by taking the action
\begin{align}
I  &  =-\frac{1}{2}%
%TCIMACRO{\dint }%
%BeginExpansion
{\displaystyle\int}
%EndExpansion
d^{4}x\sqrt{g}R+\frac{1}{2}%
%TCIMACRO{\dint }%
%BeginExpansion
{\displaystyle\int}
%EndExpansion
d^{4}x\lambda\left(  \sqrt{g}-\frac{1}{3!}\epsilon^{\mu\nu\kappa\lambda
}\partial_{\mu}X\epsilon_{abcd}Y^{a}\partial_{\nu}Y^{b}\partial_{\kappa}%
Y^{c}\partial_{\lambda}Y^{d}\right) \nonumber\\
&  +\frac{1}{2}%
%TCIMACRO{\dint }%
%BeginExpansion
{\displaystyle\int}
%EndExpansion
d^{4}x\sqrt{g}\lambda^{\prime}\left(  Y^{a}Y^{a}-1\right)  +\frac{1}{2}%
%TCIMACRO{\dint }%
%BeginExpansion
{\displaystyle\int}
%EndExpansion
d^{4}x\sqrt{g}\lambda^{^{\prime\prime}}\left(  g^{\mu\nu}\partial_{\mu
}X\partial_{\nu}X-1\right)  .
\end{align}
We have included a constraint on the $X$ field, which is known to have the
effect of replacing the scale factor in gravity by the field $X$ which mimics
dark matter \cite{CM},\cite{CMV}. We also have the option of not including
this field, and in that case the effects of the field $X$ will only be
topological providing only the joining of the disconnected pieces. For
simplicity, we have included only the coordinates of one of the maps $Y^{a}.$
First, we have the $\lambda^{\prime\prime}$ and $g^{\mu\nu}$ equations%
\begin{align}
g^{\mu\nu}\partial_{\mu}X\partial_{\nu}X  &  =1\label{constraint}\\
G_{\mu\nu}+\frac{1}{2}\lambda g_{\mu\nu}-\lambda^{^{\prime\prime}}%
\partial_{\mu}X\partial_{\nu}X  &  =0.
\end{align}
Taking the trace of Einstein equation gives
\begin{equation}
\lambda^{^{\prime\prime}}=G+2\lambda,
\end{equation}
resulting in the traceless equation%
\begin{equation}
G_{\mu\nu}-G\partial_{\mu}X\partial_{\nu}X+\frac{1}{2}\lambda\left(  g_{\mu
\nu}-4\partial_{\mu}X\partial_{\nu}X\right)  =0.
\end{equation}
Next the variation of the field $X$ gives%
\begin{equation}
\partial_{\mu}\left(  \sqrt{g}g^{\mu\nu}\partial_{\nu}X\left(  G+2\lambda
\right)  \right)  =\frac{1}{2}\partial_{\mu}\lambda V^{\mu},
\end{equation}
where we have denoted
\begin{equation}
V^{\mu}=\frac{1}{3!}\epsilon^{\mu\nu\kappa\lambda}\epsilon_{abcd}Y^{a}%
\partial_{\nu}Y^{b}\partial_{\kappa}Y^{c}\partial_{\lambda}Y^{d},
\end{equation}
and used the property%
\begin{equation}
\partial_{\mu}V^{\mu}=0.
\end{equation}
This last equation is a consequence of the identity $dY^{1}\wedge dY^{2}\wedge
dY^{3}\wedge dY^{4}=0$ which follows from%
\begin{equation}
dY^{4}=-\frac{1}{Y^{4}}\left(  Y^{1}dY^{1}+Y^{2}dY^{2}+Y^{3}dY^{3}\right)  .
\end{equation}
The $Y^{a}$ equation gives%
\begin{align}
&  \sqrt{g}\lambda^{\prime}Y_{a}-\frac{1}{2}\lambda\partial_{\mu}X\frac{1}%
{3!}\epsilon^{\mu\nu\kappa\lambda}\epsilon_{abcd}\partial_{\nu}Y^{b}%
\partial_{\kappa}Y^{c}\partial_{\lambda}Y^{d}\\
&  =\partial_{\mu}X\partial_{\nu}\left(  \lambda\frac{1}{3!}\epsilon^{\mu
\nu\kappa\lambda}\epsilon_{abcd}Y^{b}\partial_{\kappa}Y^{c}\partial_{\lambda
}Y^{d}\right)  .
\end{align}
Contracting this equation with $Y^{a}$ gives%
\begin{equation}
\lambda^{\prime}=\frac{3}{2}\lambda.
\end{equation}
The Bianchi identity gives
\begin{equation}
\frac{1}{2}\partial_{\mu}\lambda=\nabla^{\nu}\left(  \left(  G+2\lambda
\right)  \partial_{\mu}X\partial_{\nu}X\right)  . \label{bianchi}%
\end{equation}
Using the property $\left(  \nabla^{\nu}\partial_{\mu}X\right)  \partial_{\nu
}X=0$, obtained by differentiating equation (\ref{constraint}) this simplifies
to
\begin{equation}
\frac{1}{2}\partial_{\mu}\lambda=\frac{1}{\sqrt{g}}\partial_{\rho}\left(
\sqrt{g}g^{\rho\nu}\left(  G+2\lambda\right)  \right)  \partial_{\mu}X.
\end{equation}
For example, in the synchronous gauge where $g_{tt}=1$ and $X=t,$ we find
$\partial_{i}\lambda=0$ and
\begin{equation}
\frac{\partial}{\partial t}\left(  G+\frac{3}{2}\lambda\right)  +\frac{1}%
{2}\frac{\partial}{\partial t}\ln g=0.
\end{equation}
For Friedmann type universe this condition simplifies to
\begin{equation}
\frac{\partial}{\partial t}\left(  G+3\frac{\overset{\cdot}{a}}{a}+\frac{3}%
{2}\lambda\right)  =0,
\end{equation}
which is the Einstein equation allowing mimetic dark matter and cosmological
constant arising as integration constants.

One can easily verify that the Bianchi identity (\ref{bianchi}) upon
contracting by $V^{\mu}$ gives%
\begin{align}
\frac{1}{2}\partial_{\mu}\lambda V^{\mu}  &  =\partial_{\mu}XV^{\mu}%
\,\nabla^{\nu}\left(  \left(  G+2\lambda\right)  \partial_{\nu}X\right) \\
&  =\partial_{\mu}\left(  \sqrt{g}g^{\mu\nu}\left(  G+2\lambda\right)
\partial_{\nu}X\right)  ,
\end{align}
which coincides with the $X$ equation after contracting with $V^{\mu}.$

Note that if the constraint $\left(  g^{\mu\nu}\partial_{\mu}X\partial_{\nu
}X\right)  _{\Sigma}=1$ is only imposed on the boundary, then there will be no
need for a Lagrange multiplier and the equations do simplify to give
\begin{align}
G_{\mu\nu}-\frac{1}{4}g_{\mu\nu}G  &  =0\\
G+2\lambda &  =0\\
\partial_{\mu}\lambda &  =0\\
\lambda^{\prime}  &  =\frac{3}{2}\lambda.
\end{align}
without any new information from the $Y^{a}$ and $X$ equations.

\subsection{$\mathbb{R}^{2}\times S^{2}$ case}

We start with the action%
\begin{align}
I  &  =-\frac{1}{2\kappa^{2}}%
%TCIMACRO{\dint }%
%BeginExpansion
{\displaystyle\int}
%EndExpansion
d^{4}x\sqrt{g}R+\frac{1}{2}%
%TCIMACRO{\dint }%
%BeginExpansion
{\displaystyle\int}
%EndExpansion
d^{4}x\lambda\left(  \frac{1}{\kappa^{3}}\sqrt{g}-\frac{1}{2!}\epsilon^{\mu
\nu\kappa\lambda}\epsilon_{ab}\partial_{\mu}X^{a}\partial_{\nu}X^{b}%
\epsilon_{mnp}Y^{m}\partial_{\kappa}Y^{n}\partial_{\lambda}Y^{p}\right)
\nonumber\\
&  +\frac{1}{2\kappa^{4}}%
%TCIMACRO{\dint }%
%BeginExpansion
{\displaystyle\int}
%EndExpansion
d^{4}x\sqrt{g}\lambda^{\prime}\left(  Y^{m}Y^{m}-1\right)  .
\end{align}
Varying $g^{\mu\nu}$ and setting $\kappa^{2}=1,$ gives
\begin{equation}
G_{\mu\nu}+\frac{1}{2}\lambda g_{\mu\nu}=0,
\end{equation}
and by tracing this equation we get%
\begin{equation}
G+2\lambda=0.
\end{equation}
After substituting back we get the traceless equation%
\begin{equation}
G_{\mu\nu}-G\partial_{\mu}X\partial_{\nu}X+\frac{1}{2}\lambda g_{\mu\nu}=0.
\end{equation}
The Bianchi identity gives%
\begin{equation}
\frac{1}{2}\partial_{\mu}\lambda=0.\nonumber
\end{equation}
Next, we have the $X^{a}$ equation
\begin{align}
&  -\partial_{\mu}\left(  \epsilon^{\mu\nu\kappa\lambda}\epsilon_{ab}%
\lambda\partial_{\nu}X^{b}\epsilon_{mnp}Y^{m}\partial_{\kappa}Y^{n}%
\partial_{\lambda}Y^{p}\right) \label{xequation}\\
&  =\partial_{\mu}\lambda\epsilon^{\mu\nu\kappa\lambda}\epsilon_{ab}%
\partial_{\nu}X^{b}\epsilon_{mnp}Y^{m}\partial_{\kappa}Y^{n}\partial_{\lambda
}Y^{p},\nonumber
\end{align}
and finally the $Y^{m}$ equation gives
\begin{align}
&  \sqrt{g}\lambda^{\prime}Y_{m}-\frac{1}{2}\lambda\epsilon_{ab}\partial_{\mu
}X^{a}\partial_{\nu}X^{b}\frac{1}{2}\epsilon^{\mu\nu\kappa\lambda}%
\epsilon_{mnp}\partial_{\kappa}Y^{n}\partial_{\lambda}Y^{p}\\
&  =\epsilon_{ab}\partial_{\mu}X^{a}\partial_{\nu}X^{b}\partial_{\kappa
}\left(  \lambda\frac{1}{2}\epsilon^{\mu\nu\kappa\lambda}\epsilon_{mnp}%
Y^{n}\partial_{\lambda}Y^{p}\right)  .
\end{align}
Contracting this equation with $Y^{m}$ gives%
\begin{equation}
\lambda^{\prime}=\frac{3}{2}\lambda,
\end{equation}
and thus
\begin{equation}
\frac{3}{2}\sqrt{g}Y_{m}-\frac{3}{2}\epsilon_{ab}\partial_{\mu}X^{a}%
\partial_{\nu}X^{b}\frac{1}{2}\epsilon^{\mu\nu\kappa\lambda}\epsilon
_{mnp}\partial_{\kappa}Y^{n}\partial_{\lambda}Y^{p}=0,
\end{equation}
together with
\begin{equation}
\sqrt{g}=\frac{1}{2!}\epsilon^{\mu\nu\kappa\lambda}\epsilon_{ab}\partial_{\mu
}X^{a}\partial_{\nu}X^{b}\epsilon_{mnp}Y^{m}\partial_{\kappa}Y^{n}%
\partial_{\lambda}Y^{p}.
\end{equation}
This implies%
\begin{equation}
Y_{m}\epsilon_{pqr}Y^{p}\partial_{\left[  \kappa\right.  }Y^{q}\partial
_{\left.  \lambda\right]  }Y^{r}=\epsilon_{mnp}\partial_{\left[
\kappa\right.  }Y^{n}\partial_{\left.  \lambda\right]  }Y^{p}.
\end{equation}
This relation is an identity which follows from the vanishing of a rank four
antiysmmetric tensor $\left[  mpqr\right]  $ taking three values
\begin{align}
0  &  =Y_{\left[  m\right.  }\epsilon_{pq\left.  r\right]  }\partial_{\left[
\kappa\right.  }Y^{q}\partial_{\left.  \lambda\right]  }Y^{r}\\
&  =Y_{m}\epsilon_{pqr}Y^{p}\partial_{\left[  \kappa\right.  }Y^{q}%
\partial_{\left.  \lambda\right]  }Y^{r}-Y_{p}\epsilon_{mqr}\partial_{\left[
\kappa\right.  }Y^{q}\partial_{\left.  \lambda\right]  }Y^{r}+2Y_{q}%
\epsilon_{pmr}Y^{p}\partial_{\left[  \kappa\right.  }Y^{q}\partial_{\left.
\lambda\right]  }Y^{r},
\end{align}
the last term being zero because $Y_{q}\partial_{\kappa}Y^{q}=0.$ Thus, as
expected, no new information comes from the $Y$ equation, except for its trace.

The $X_{a}$ equation reduces to
\begin{align}
&  -\partial_{\mu}\left(  \epsilon^{\mu\nu\kappa\lambda}\epsilon_{ab}%
\lambda\partial_{\nu}X^{b}\epsilon_{mnp}Y^{m}\partial_{\kappa}Y^{n}%
\partial_{\lambda}Y^{p}\right) \nonumber\\
&  =\partial_{\mu}\lambda\epsilon^{\mu\nu\kappa\lambda}\epsilon_{ab}%
\partial_{\nu}X^{b}\epsilon_{mnp}Y^{m}\partial_{\kappa}Y^{n}\partial_{\lambda
}Y^{p},
\end{align}
which is identically satisfied since $\partial_{\mu}\lambda=0.$ This shows
that the resulting system is that of gravity plus mimetic dark matter, with
the topological fields $Y^{m}\left(  x\right)  $ connecting the different unit
spheres, constituting the building fabric of space-time.

Finally we comment on the possibility of adding mimetic matter to the system
corresponding to the quantization of $\mathbb{R}^{2}\times\Sigma_{2}$ where
$\Sigma_{2}$ is a two dimensional surface. Looking at the induced metric
\begin{equation}
h^{ab}=g^{\mu\nu}\partial_{\mu}X^{a}\partial_{\nu}X^{b},\qquad a,b=1,2,
\end{equation}
we notice that we had to impose, on the boundary, the constraint
\begin{equation}
\det h^{ab}=1,
\end{equation}
which is the area preserving condition for the two dimensional surfaces. These
maps will be characterized by the value of the trace of $h^{ab}$ and their
stability will depend on the value of $t=\mathrm{tr}$ $h^{ab}.$ These are
stable and of the elliptic type when $-2<t<2.$ Unfortunately, the resulting
system of equations is not easy to solve, and it is not clear whether such
system can lead to realistic models. It is therefore doubtful whether using
more than one scalar field associated with imposing one or more constraints is
useful. We conclude that for our purposes, it is enough to characterize the
conditions for area quantization is to have an area preserving conditions on
the mapping defined by the two fields $X^{1}$ and $X^{2}$ taken as boundary conditions.

\section{Discussion and conclusions}

It is an ambitious goal to initiate a program of axiomatization of physics as
suggested by Hilbert. Our proposal is to start from an analogue of the
Heisenberg commutation relation to quantize the geometry. The Dirac operator
plays the role of momentum while the Feynman slash of scalar fields plays the
role of coordinates. When the dimension of the noncommutative space, as
determined by the growth of eigenvalues, is $2$ or $4$ there are two possible
Clifford algebras with which the scalar fields are contracted with the
corresponding gamma matrices. These two Clifford algebras are related to each
other through the reality operator $J$ $\ $which is an anti-unitary operator
that is part of the data defining the noncommutative space. In four dimensions
the sum of the two Clifford algebras is $M_{2}\left(  \mathbb{H}\right)
\oplus M_{4}\left(  \mathbb{C}\right)  $ which is the algebra of the finite
space that is tensored with the continuous Riemannian space. The quantization
condition implies that the volume of the continuos part of the space is
quantized in terms of the winding numbers of the two mappings $Y$ and
$Y^{\prime}$ from $M_{4}$ to $S^{4}.$ The presence of two maps instead of one
allows for the representation of a spin-manifold $M_{4},$ with arbitrary
topology and large volume as the pullback of the two maps which yields four
coordinates given on local charts. This construction determines, in a unique
way, the noncommutative space that defines our space-time. Inner fluctuations
of the Dirac operator by automorphisms of the algebra extends it to include a
connection, which is a one form defined over the noncommutative space.
Components of the connection along the continuous directions are the gauge
fields of the resulting gauge group, and the components along the discrete
directions are the Higgs fields. The connection then includes all the bosonic
fields of a unified field theory, which is a Pati-Salam model with a definite
Higgs structure. There are two special cases when these Higgs fields are
either truncated or are in composite representations of more fundamental
fields. The Standard Model with neutrinos (and a singlet) is a special case of
the Pati-Salam model which satisfies an order one condition where the
connection becomes restricted to the algebra $\mathcal{A}$ but not its
opposite. Elements of the Hilbert space define the fermions which are $16$ in
the representation $\left(  2,1,4\right)  +\left(  1,2,4\right)  $ with
respect to the symmetry $S\left(  2\right)  _{R}\times SU\left(  2\right)
_{L}\times SU\left(  3\right)  .$ Thus all bosonic fields in the -form of
gravity, gauge and Higgs fields are unified in the Dirac operator and all
fermion fields are unified in the fundamental representation in the Hilbert
space. \ The dynamics is governed by the spectral action principle where the
spectral action is an arbitrary positive function of the Dirac operator valid
up to a cutoff scale, which is taken to be near the Planck scale. In other
words, by starting from a quantization condition on the volume of the
noncommutative space, all fields and their interactions are predicted and
given by a Pati-Salam model which has three special cases one of which is the
Standard Model with neutrino masses and a singlet field. The spectral Standard
Model predicts unification of gauge couplings and the correct mass for the top
quark and is consistent with a low Higgs mass of $125$ Gev. The unification
model is assumed to hold at the unification scale and when the gauge, Yukawa
and Higgs couplings relations are taken as initial conditions on the RGE, one
finds complete agreement with experiment, except for the meeting of the gauge
couplings which are off by $4\%.$ This suggests that a Pati-Salam model
defines the physics beyond the Standard Model, and where we have shown
\cite{Walter} that it allows for unification of gauge couplings, consistent
with experimental data.

The assumption of volume quantization has consequences on the structure of
General Relativity. Equations of motion agree with Einstein equations except
for the trace condition, which now determines the Lagrange multiplier
enforcing volume quantization. The cosmological constant, although not
included in the action, is now an integration constant. The two mapping fields
$Y$ and $Y^{\prime}$ from the four-manifold to $S^{4}$ can be considered to be
be solutions of instanton equations and give the physical picture that
coordinates of a point are represented as the localization of instantons with
finite energy. To have a physical picture of time we have also considered a
four-manifold formed with the topology of $R\times\Sigma_{3}$, where
$\Sigma_{3}$ is a three dimensional hypersurface, to allow for space-times
with Lorentzian signature. The quantization condition is modified to have two
mappings from $\Sigma_{3}\rightarrow S^{3}$ and a mapping $X:\mathbb{R}%
\rightarrow\mathbb{R}.$ The resulting algebra of the noncommutative space is
unchanged, and the three dimensional volume is quantized provided that the
mapping field $X$ is constrained to have unit gradient. This field $X$
modifies only the longitudinal part of the graviton and plays the role of
mimetic dust. It thus solves, without extra cost, the dark matter problem
\cite{CM}. Recently, we have shown that this field $X$ can be used to build
realistic cosmological models \cite{CMV}. In addition, and under certain
conditions, could be used to avoid singularities in General relativity for
Friedmann, Kasner \cite{CMK} and Black hole solutions \cite{CMBH}. This is
possible because this scalar field modifies the longitudinal sector in GR. We
have presented various implications of the quantization condition such as the
absence of the cosmological constant from the action, quantizing volumes and
areas of maps of $M_{4}$ to $S^{4}$, $\mathbb{R}\times S^{3}$ and
$\mathbb{R}^{2}\times S^{2}.$

We have presented enough evidence that a framework where space-time assumed to
be governed by noncommutative geometry results in a unified picture of all
particles and their interactions. The axioms could be minimized by starting
with a volume quantization condition, which is the Chern character formula of
the noncommutative space and a special case of the orientability condition.
This condition determines uniquely the structure of the noncommutative space.
Remarkably, the same structure was also derived, in slightly less unique way,
by classifying all finite noncommutative spaces \cite{CC2}. The picture is
very compelling, in contrast to other constructions, such as grand
unification, supersymmetry or string theory, where there is no limit on the
number of possible models that could be constructed. The picture, however, is
still incomplete as there are still many unanswered questions and we now list
few of them. Further studies are needed to determine the structure and
hierarchy of the Yukawa couplings, the number of generations, the form of the
spectral function and the physics at unification scale, quantizing the fields
appearing in the spectral action and in particular the gravitational field. To
conclude, noncommutative geometry as a basis for unification, is a predictive
and exciting field with very appealing features and many promising new
directions for research.

\begin{acknowledgement}
I would like to thank Alain Connes for a fruitful and pleasant collaboration
on the topic of noncommutative geometry for the last twenty years. I would
also like to thank Walter van Suijlekom and Slava Mukhanov for essential
contributions to this program of research. This research is supported in part
by the National Science Foundation under Grant No. Phys-1518371.
\end{acknowledgement}

\end{document}